\documentclass[utf8]{FrontiersinHarvard} 
\usepackage{url,lineno}
\usepackage[onehalfspacing]{setspace}

\usepackage{graphicx}				
\usepackage{amssymb}
\usepackage{comment}
\usepackage{tgschola}
\usepackage[mathscr]{euscript}
\usepackage{color}
\renewcommand{\baselinestretch}{1.15}
\usepackage{mathrsfs}
\usepackage{lipsum}
\usepackage{hyperref}
\usepackage{xcolor}
\definecolor{xll}{rgb}{0.85,0.44,0.84}
\hypersetup{
  colorlinks=true,
  citecolor=xll,
  linkcolor=cyan,
  urlcolor=magenta,
}

\usepackage{amsmath,amsfonts,amssymb}
\usepackage{mathtools}

\usepackage{multicol}
\usepackage[T1]{fontenc}
\usepackage{epstopdf}
\usepackage{empheq}

\usepackage{titlesec}
\renewcommand{\thesection}{\Roman{section}}
\usepackage{pxfonts}
\titleformat{\section}{\Large\helveticabold}{\thesection}{.5em}{}
\usepackage{fouriernc}

\DeclareMathAlphabet{\mathcal}{OMS}{cmsy}{m}{n}
\renewcommand{\d}{\mathrm{d}}

\renewcommand{\bf}[1]{\mathbf{#1}}
\renewcommand{\rm}[1]{\mathrm{#1}}

\usepackage[symbol]{footmisc}
\numberwithin{equation}{section}

\usepackage{chngcntr}
\counterwithout{equation}{section}
\renewcommand\theequation{\arabic{equation}}
\renewcommand\thefigure{\arabic{figure}}
\usepackage{wrapfig}

\def\thesection{\arabic{section}\hspace*{0.cm}}

\usepackage{authblk}

\usepackage{hyperref}
\usepackage{cleveref}

\renewenvironment{abstract}
 {\normalsize
 {\hspace*{-0.4cm}
 \sc\sf\bfseries\helveticabold \abstractname\vspace{-0.2cm}\vspace{0pt}
 }
  \list{}{
    \setlength{\leftmargin}{0mm}
    \setlength{\rightmargin}{\leftmargin}
  }
  \item\relax}
 {\endlist}

\usepackage{indentfirst}
\newcommand{\x}{\mathrm{X}}

\newcommand{\epsc}{\varepsilon_{\rm{c}}}
\newcommand{\Pc}{P_{\rm{c}}}
\newcommand{\hP}{\widehat{P}}
\newcommand{\hr}{\widehat{r}}
\newcommand{\hM}{\widehat{M}}
\newcommand{\heps}{\widehat{\varepsilon}}

\nolinenumbers
\def\keyFont{\fontsize{9}{11}\helveticabold }

\def\firstAuthorLast{Cai and Li} 

\def\Authors{{\helveticabold Bao-Jun Cai$^{1,*}$}, {\helveticabold Bao-An Li\,$^{2,*}$}}

\def\Address{\small$^{1}$Quantum Machine Learning Laboratory, Shadow Creator Inc., Shanghai 201208,
 People's Republic of China \\
$^{2}$Department of Physics and Astronomy, Texas A$\&$M University-Commerce,
 Commerce, TX 75429-3011, USA}

\def\corrAuthor{\scriptsize Bao-An Li}
\def\corrEmail{\scriptsize Bao-An.Li@tamuc.edu}

\usepackage{natbib}

\begin{document}
\onecolumn
\firstpage{1}

\title[New Insights into Cold Supradense Matter in Neutron Star Cores]{New Insights into Supradense Matter
from Dissecting Scaled Stellar Structure Equations} 

\author[\firstAuthorLast ]{\Authors} 
\address{} 
\correspondance{\scriptsize \selectfont\fontfamily{lmr}} 

\extraAuth{}

\maketitle
\begin{abstract}
{\small
 The strong-field gravity in General Relativity (GR) realized in neutron stars (NSs) renders the Equation of State (EOS) $P(\varepsilon)$ of supradense neutron star (NS) matter to be essentially nonlinear and refines the upper bound for $\phi\equiv P/\varepsilon$ to be much smaller than the Special Relativity (SR) requirement with linear EOSs, where $P$ and $\varepsilon$ are respectively the pressure and energy density of the system considered.
Specifically, a tight bound $\phi\lesssim0.374$ is obtained by anatomizing perturbatively the intrinsic structures of the scaled Tolman--Oppenheimer--Volkoff (TOV) equations without using any input nuclear EOS. New insights gained from this novel analysis provide EOS-model independent constraints on properties (e.g., density profiles of the sound speed squared $s^2=\d P/\d\varepsilon$ and trace anomaly $\Delta=1/3-\phi$) of cold supradense matter in NS cores. Using the gravity-matter duality in theories describing NSs, we investigate the impact of gravity on supradense matter EOS in NSs. In particular, we show that the NS mass $M_{\rm{NS}}$, radius $R$ and its compactness $\xi\equiv M_{\rm{NS}}/R$ scale with certain combinations of its central pressure and energy density (encapsulating its central EOS). Thus, observational data on these properties of NSs can straightforwardly constrain NS central EOSs without relying on any specific nuclear EOS-model.
\\\\
\keyFont{\bfseries Keywords: Equation of State,  Supradense Matter,  Neutron Star, Tolman--Oppenheimer--Volkoff Equations, Principle of Causality, Special Relativity, Speed of Sound, Generality Relativity, Neutron-rich Matter, Gravity-matter Duality}
}\end{abstract}

\section{Introduction}\label{SEC_1}

\begin{figure}[ht!]
\centering
\includegraphics[width=7.cm]{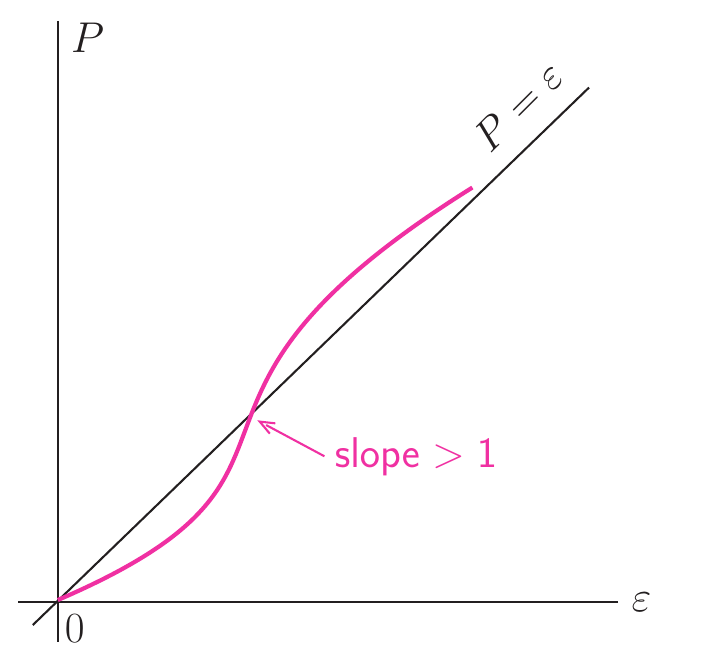}
\caption{\small (Color Online). An illustration of the equivalence between $s^2\leq1$ with a linear EOS and $\phi\leq1$: If $P$ could be greater than $\varepsilon$ somewhere then the curve of $P(\varepsilon)$ have to across the line $P=\varepsilon$ from below to above, indicating that $s^2=\d P/\d\varepsilon>1$ at the crossing point.}\label{fig_SRu}
\end{figure}

The speed of sound squared (SSS) $s^2=\d P/\d\varepsilon$\,\citep{Landau1987} quantifies the stiffness of the Equation of State (EOS) expressed in terms of the relationship $P(\varepsilon)$ between the pressure $P$ and energy density $\varepsilon$ of the system considered. The Principle of Causality of Special Relativity (SR) requires the speed of sound of any signal to stay smaller than the speed of light $c\equiv 1$, i.e., $s\leq1$. For a linear EOS of the form $P=w\varepsilon$ with $w$ being some constant, 
the condition $s^2\leq1$ is globally equivalent to $\phi=P/\varepsilon\leq1$. For such EOSs, the causality condition can be equivalently written as:
\begin{equation}\label{SR-1}
\boxed{
\mbox{Principle of Causality of SR with linear EOS implies }P\leq\varepsilon\leftrightarrow\phi\equiv P/\varepsilon\leq1.}
\end{equation}
The indicated equivalence between $s^2\leq1$ and $\phi\leq1$ could be demonstrated straightforwardly as follows: If $P$ could be greater than $\varepsilon$ somewhere then the curve of $P(\varepsilon)$ may unavoidably across the line $P=\varepsilon$ from below to above, indicating the slope at the crossing point is necessarily larger than 1, as illustrated in FIG.\,\ref{fig_SRu}. In the following, we use the above causality requirement on $\phi$ with linear EOSs as a reference in discussing properties of supradense matter in strong-field gravity.

\renewcommand*\figurename{\small FIG.}

The EOS of nuclear matter may be strongly nonlinear depending on both the internal interactions and the external environment/constraint of the system; this means that $\phi\leq1$ is necessary but not sufficient to ensure supradense matter in all NSs always stay casual. For example, the EOS of noninteracting degenerate Fermions (e.g., electrons) can be written in the polytropic form $P=K\varepsilon^{\beta}$\,\citep{Shapiro1983} where $\beta=5/3$ for non-relativistic and $\beta=4/3$ for extremely relativistic electrons; consequently $\phi\leq\beta^{-1}<1$.
Similarly, a long time ago Zel'dovich considered the EOS of an isolated ultra-dense system of baryons interacting through a vector field\,\citep{Zeldovich61}. In this case, $P=\varepsilon\sim\rho^2$, here $\rho$ is the baryon number density. Consequently, $P/\varepsilon\leq1$ is obtained. The EOS of dense nuclear matter where nucleons interact through both the $\sigma$-meson and $\omega$-meson in the Walecka model\,\citep{Walecka1974} is an example of this type.
In particular, the $\omega$-field scales at asymptotically large density as $\omega\sim\rho$ while the $\sigma$-field scales $\sigma\sim\rho_{\rm{s}}$ with the scalar density $\rho_{\rm{s}}$ approaching some constant for $\rho\to\infty$\,\citep{CaiLi2016}; therefore the vector field dominates at these densities. More generally, however, going beyond the vector field, the baryon density dependence of either $P(\rho)$ or $\varepsilon(\rho)$ could be very complicated and nontrivial. The resulting EOS $P(\varepsilon)$ could also be significantly nonlinear.
The EOS of supradense matter under intense gravity of NSs could be forced to be nonlinear as the equilibrium state of NSs is determined by extremizing the total action of the matter-gravity system through the Hamilton's variational principle. It is well known that the strong-field gravity in General Relativity (GR) is fundamentally nonlinear, the EOS of NS matter especially in its core is thus also expected to be nonlinear. Therefore, the causality condition $s^2\leq1$ may be appreciably different from $\phi\leq1$, and it may also effectively render the upper bound for $\phi$ to be smaller than 1.
Determining accurately an upper bound of $\phi$ (equivalently a lower bound of the dimensionless trace anomaly $\Delta=1/3-\phi$) will thus help constrain properties of supradense matter in strong-field gravity.

The upper bound for $\phi$ is a fundamental quantity encapsulating essentially the strong-field properties of gravity in GR. Its accurate determination may help improve our understanding about the nature of gravity. The latter is presently least known among the four fundamental forces despite being the first one discovered in Nature\,\citep{Hoyle:2003dw}.
An upper bound on $\phi$ substantially different from 1 then vividly characterizes how GR affects the supradense matter existing in NSs.
In some physical senses, this is similar to the effort in determining the Bertsch parameter. The latter was introduced as the ratio $E_{\rm{UFG}}/E_{\rm{FFG}}$ of the EOS of a unitary Fermi gas (UFG) over that of the free Fermi gas (FFG) $E_{\rm{FFG}}$\,\citep{Bloch2008};  here $E_{\rm{FFG}}$ and $E_{\rm{UFG}}$ are the energies per particle in the two systems considered. It characterizes the strong interactions among Fermions under the unitary condition. Extensive theoretical and experimental efforts have been made to constrain/fix the Bertsch parameter. Indeed, its accurate determination has already made strong impact on understanding strongly-interacting Fermions\,\citep{Bloch2008,Gior2008}.

There are fundamental physics issues regarding both strong-field gravity and supradense matter EOS as well as their couplings. What is gravity?  Is a new theory of light and matter needed to explain what happens at very high energies and temperatures? These are among the eleven greatest unanswered physics questions for this century
identified in 2003 by the National Research Council of the US National Academies\,\citep{11questions}. Compact stars provide far more extreme conditions necessary to test possible answers to these questions
than terrestrial laboratories. A gravity-matter duality exists in theories describing NS properties, see, e.g., Refs.\,\citep{Psaltis:2008bb,Shao:2019gjj} for recent reviews. Neutron stars are natural testing grounds of our knowledge on these issues. 
Some of their observational properties may help break the gravity-matter duality, see, e.g., Refs.\,\citep{DeDeo:2003ju,Wen:2009av,Lin:2013rea,He:2014yqa,Yang:2019rxn}. Naturally, these issues are intertwined and one may gain new insights about the EOS of supradense matter from analyzing features of strong-field gravity or vice versa. 
The matter-gravity duality reflects the deep connection between microscopic physics of supradense matter and the powerful gravity effects of NSs. They both have to be fully understand to unravel mysteries associated with compact objects in the Universe. In this brief review, we summarize the main physics motivation, formalism and results of our recent efforts to gain new insights into the EOS of supradense matter in NS cores by dissecting perturbatively the intrinsic structures of the Tolman--Oppehnheimer--Volkoff (TOV) equations\,\citep{TOV39-1,TOV39-2} without using any input nuclear EOS. For more details, we refer the readers to our original publications in Refs.\,\citep{CLZ23-a,CLZ23-b,CL24-a,CL24-b}.

The rest of this paper is organized as follows: 
First of all, to be complete and easy of our following presentation, in Section \ref{remaks} we make a few remarks about some existing constraints on the EOS of supradense NS matter. 
Section \ref{SEC_2} introduces the scaled TOV equations starting from which one can execute an effective perturbative expansion; the central SSS is obtained in Section \ref{SEC_3}, we then infer an upper bound for the ratio $\x\equiv\phi_{\rm{c}}=P_{\rm{c}}/\varepsilon_{\rm{c}}$ of central pressure $P_{\rm{c}}$ over central energy density $\varepsilon_{\rm{c}}$ for NSs at the maximum-mass configuration along the M-R curve.
The generalization for the upper bound of $P/\varepsilon$ is also studied in Section \ref{SEC_3}. In Section \ref{SEC_DELTA}, we compare our prediction on the lower bound of $\Delta=1/3-P/\varepsilon$ with existing predictions in the literature. We summarize in Section \ref{SEC_4} and give some perspectives for future studies along this line.
In the Appendix, we discuss an effective correction to $s_{\rm{c}}^2$ obtained in Section \ref{SEC_3}.

\section{Remarks on Some Existing Constraints on Supradense NS Matter}\label{remaks}
Understanding the EOS of supradense matter has long been an important issue in both nuclear physics and astrophysics\,\citep{Walecka1974,Chin1977,Freedman1977-3,Baluni1978,Wiringa1988,Akmal1998,Migdal1978,Morley1979,Shuryak1980,Bailin1984,Lattimer2001,Dan02,Steiner2005,Lattimer2007PR,
Alford2008,LCK08,Watts2016,Ozel2016,Oertel2017,Vidana2018}.
In fact, it has been an outstanding science driver 
at many research facilities in both fields. 
For example, finding the EOS of densest visible matter existing in our Universe is an ultimate goal of astrophysics in the era of high-precision multimessenger astronomy\,\citep{sathyaprakash2019}.
However, despite of much effort and progress made during the last few decades using various observational data and models especially since the discovery of GW170817\,\citep{Abbott2017,Abbott2018}, GW190425\,\citep{Abbott2020-a}, GW190814\,\citep{Abbott2020} and the recent NASA's NICER (Neutron Star Interior Composition Explorer) mass-radius measurements for PSR J0740+6620\,\citep{Fon21,Riley21,Miller21,Salmi22,Ditt24,Salmi24},  PSR J0030+0451\,\citep{Riley19,Miller19,Vin24} and PSR J0437-4715\,\citep{Choud24,Reard24}, knowledge about NS core EOS remains ambiguous and quite elusive, see, e.g., Refs.\,\citep{Bose2018,De2018,Fatt2018,Lim2018,Most2018,Radice2018,Tews2018,ZhangLiXu2018,Baus2019,Baus2020,Baym2019,McLerran2019,Most2019,Ann20,Ann23,Sed2020,Zhao2020,Weih2020,Xie2019,XieLi2020,Xie2021,Drischler2020,Drischler2021,Li:2020dst,Bombaci2021,Mam2021,Nathanail2021,Raaij2021,Altiparmak2022,Breschi2022,Kom2022,Perego2022,Huang2022,Tan2022-a,Tan2022-b,Brandes2023,Brandes2023-a,Gorda2023,Han2023,Jiang2023,Ofeng2023,Mro23,Raithel2023,Som2023,ZhangLi2020,ZhangLi2021x,ZhangLi2023b,ZhangLi2023a,Pang2023,Fuji2024b,Pro2024,Ruther2024}. For more discussions, see recent reviews, e.g., Refs.\,\citep{Baym2018,Bai2019,Li2019,Ors2019,Blaschke:2020qqj,Capano:2019eae,Chatziioannou:2020pqz,Bur2021,Dex2021,Dri2021,Lattimer:2021emm,LCXZ2021,Lov2022,Sedr23,Kumar2024,Sor2024,Tsang:2023vhh}.

Extensive theoretical investigations about the EOS of supradense NS matter have been done and many interesting predictions have been made. 
For example, the realization of approximate conformal symmetry of quark matter at extremely high densities $\rho\gtrsim40\rho_0$ with $\rho_0\equiv\rho_{\rm{sat}}$ the nuclear saturation density implies the corresponding EOS approaches that of an ultra-relativistic Fermi gas (URFG) from below, namely\,\citep{Bjorken83,Kur10}
\begin{equation}
\boxed{
\rm{URFG:}~~    P\lesssim{\varepsilon}/{3}\leftrightarrow\phi\lesssim1/3,~~\mbox{at extremely high densities}.}
\end{equation}
For the URFG, $3P\approx\varepsilon\sim\rho^{4/3}$.
Therefore $\phi=P/\varepsilon$ is at least upper bounded to be below 1/3 at these densities, equivalently a lower bound on the dimensionless trace anomaly emerges:
\begin{equation}
\boxed{
    \Delta\equiv 1/3-P/\varepsilon\gtrsim0,~~\mbox{at extremely high densities }\rho\gtrsim40\rho_0.}
\end{equation}
This prompts the question whether the bound $\phi\leq1/3$ holds globally for dense matter or some other bound(s) on $\phi$ may exist. In this sense, massive NSs like PSR J1614-2230\,\citep{Dem10,Arz18}, PSR J0348+0432\,\citep{Ant13}, PSR J0740+6620\,\citep{Fon21,Riley21,Miller21,Salmi22,Ditt24,Salmi24}
and PSR J2215+5135\,\citep{Sul24} provide an ideal testing bed for exploring such quantity.
A sizable $\phi\gtrsim\mathcal{O}(0.1)$ arises for NSs but not for ordinary stars or low-density nuclear matter\,\citep{CL24-a}. For example, considering stars such as white dwarfs (WDs), one has $P\lesssim 10^{22\mbox{-}23}\,\rm{dynes}/\rm{cm}^2\approx10^{-(11\mbox{-}10)}\,\rm{MeV}/\rm{fm}^3$ and $\varepsilon\lesssim10^{8\mbox{-}9}\,\rm{kg}/\rm{m}^3\sim10^{-6}\,\rm{MeV}/\rm{fm}^3$, thus $\phi\lesssim10^{-(5\mbox{-}4)}$. 
The $\phi$ could be even smaller for main-sequence stars like the sun. 
Specifically, the pressure and energy density in the solar core are about $10^{-16}\,\rm{MeV}/\rm{fm}^3$ and $10^{-10}\,\rm{MeV}/\rm{fm}^3$, respectively, and therefore $\phi\approx 10^{-6}$.
These stars are Newtonian in the sense that GR effects are almost absent. Similarly, for NS matter around nuclear saturation density $\rho_0=\rho_{\rm{sat}}\approx0.16\,\rm{fm}^{-3}$, the pressure is estimated to be $
P(\rho_0)\approx P_0(\rho_0)+P_{\rm{sym}}(\rho_0)\delta^2
\approx3^{-1}L\rho_0\delta^2\lesssim3\,\rm{MeV}/\rm{fm}^3$. Its isospin-dependent part is $P_{\rm{sym}}(\rho_0)=3^{-1}L\rho_0$ with $L\approx60\,\rm{MeV}$\,\citep{Li2018PPNP,LCXZ2021} being the slope parameter of nuclear symmetry energy $E_{\rm{sym}}(\rho)$ at $\rho_0$, $\delta$ is the isospin asymmetry of the system ($\delta^2\lesssim1$), and $P_0(\rho_0)=0$ is the pressure of symmetric nuclear matter (SNM) at $\rho_0$. The energy density at $\rho_0$ is similarly estimated as $
\varepsilon(\rho_0)\approx[E_0(\rho_0)+E_{\rm{sym}}(\rho_0)\delta^2+M_{\rm{N}}]\rho_0\approx150\,\rm{MeV}/\rm{fm}^3$ with $M_{\rm{N}}\approx939\,\rm{MeV}$ the nucleon static mass, $E_0(\rho_0)\approx-16\,\rm{MeV}$ the binding energy at $\rho_0$ for SNM and $E_{\rm{sym}}(\rho_0)\approx32\,\rm{MeV}$\,\citep{Li:2017nna}, leading to $\phi\lesssim0.02$.

Based on the dimensional analysis and the definition of sound speed, we may write out the SSS generally as (we use the units in which $c=1$)
\begin{equation}\label{def-s2fphi}
    s^2=\phi f(\phi),~~\phi=P/\varepsilon,
\end{equation}
where $f(\phi)$ is dimensionless.
For low-density matter, such as those in ordinary stars and WDs or the nuclear matter around saturation density $\rho_0$, the ratio $\phi$ is also  small (as estimated in the last paragraph), indicating that $f(\phi)$ could be expanded around $\phi=0$ as $f(\phi)\approx f_0+f_1\phi+f_2\phi^2+\cdots$, where $f_0>0$ (to guarantee the stability condition $s^2\geq0$).
Keeping the first leading-order term $f_0$ enables us to obtain $s^2\approx f_0\phi$, so $s^2$ has a similar value of $\phi$ if $f_0\sim\mathcal{O}(1)$ and the EOS does not take a linear form (except for $f_0=1$). Moreover, the causality principle requires $\phi\lesssim f_0^{-1}$.
The $s^2\approx0.03\sim\phi\lesssim0.02$ at $\rho_0$ from chiral effective field calculations\,\citep{Essick2021} confirms our order-of-magnitude estimate on $s^2$.
If the next-leading-order term $f_1$ is small and positive, then the upper bound for $\phi$ becomes $\phi\lesssim f_0^{-1}(1-f_1/f_0^2)$ which is even reduced compared with $f_0^{-1}$.
The exact form of $f(\phi)$ should be worked out/analyzed by the general-relativistic structure equations for NSs\,\citep{TOV39-1,TOV39-2}.
By doing that, we demonstrated earlier that $\phi$ is upper bounded as $\phi\lesssim0.374$ near the centers of stable NSs\,\citep{CLZ23-a,CLZ23-b,CL24-a,CL24-b}. 
The corresponding trace anomaly $\Delta$ in NS cores is thus bounded to be above $-$0.04.
In the next sections, we first show the main steps leading to these conclusions and then discuss their ramifications in comparison with existing predictions on $\Delta$ in the literature.

\section{Analyzing Scaled TOV Equations, Mass/Radius Scalings and Central SSS}\label{SEC_2}

The TOV equations describe the radial evolution of pressure $P(r)$ and mass $M(r)$ of a NS under static hydrodynamic equilibrium conditions\,\citep{TOV39-1,TOV39-2}.
In particular, we have (adopting $c=1$)
\begin{align}
\frac{\d P}{\d r}=-\frac{GM\varepsilon}{r^2}\left(1+\frac{P}{\varepsilon}\right)\left(1+\frac{4\pi r^3P}{M}\right)
\left(1-\frac{2GM}{r}\right)^{-1},~~
\frac{\d M}{\d r}=4\pi r^2\varepsilon,
\end{align}
here the mass $M=M(r)$, pressure $P=P(r)$ and energy density $\varepsilon=\varepsilon(r)$ are functions of the distance $r$ from NS center.
The central energy density $\epsc$ is a specific and important quantity, which straightforwardly connects the central pressure $\Pc$ via the EOS $\Pc=P(\epsc)$.
Using $\epsc$, we can construct a mass scale $W$ and a length scale $Q$:
\begin{equation}\label{RE-WQ}
W=\frac{1}{G}\frac{1}{\sqrt{4\pi G\varepsilon_{\rm{c}}}}
=\frac{1}{\sqrt{4\pi\varepsilon_{\rm{c}}}}
,~~Q=\frac{1}{\sqrt{4\pi G\epsc}}=\frac{1}{\sqrt{4\pi\varepsilon_{\rm{c}}}},
\end{equation}
respectively, here the second relations follow with $G=1$.
Using $W$ and $Q$, we can rewrite the TOV equations in the following dimensionless form\,\citep{CLZ23-a,CLZ23-b,CL24-a,CL24-b},
\begin{equation}\label{TOV-ds}
\boxed{
\frac{\d\widehat{P}}{\d\widehat{r}}
=-\frac{\widehat{\varepsilon}\widehat{M}}{\widehat{r}^2}
\frac{(1+{\widehat{P}}/{\widehat{\varepsilon}})
(1+{\widehat{r}^3\widehat{P}}/{\widehat{M}})}{1-
{2\widehat{M}}/{\widehat{r}}},~~\frac{\d\widehat{M}}{\d\widehat{r}}=\widehat{r}^2\widehat{\varepsilon},}
\end{equation}
where $\widehat{P}=P/\epsc$, $\heps=\varepsilon/\varepsilon_{\rm{c}}$, $\widehat{r}=r/Q$ and $\hM=M/W$.
The general smallness of 
\begin{equation}
\boxed{
\x\equiv\phi_{\rm{c}}\equiv\widehat{P}_{\rm{c}}\equiv P_{\rm{c}}/\varepsilon_{\rm{c}},}
\end{equation}
together with the smallness of 
\begin{equation}
\boxed{
\mu\equiv\widehat{\varepsilon}-\widehat{\varepsilon}_{\rm{c}}=\widehat{\varepsilon}-1,}
\end{equation}
near NS centers enable us to develop effective/controllable expansion of a relevant quantity $\mathcal{U}$ over $\x$ and $\mu$ as\,\citep{CLZ23-a,CLZ23-b,CL24-a,CL24-b}
\begin{equation}
\boxed{
\mathcal{U}/\mathcal{U}_{\rm{c}}\approx1+\sum_{i+j\geq1}u_{ij}\x^i\mu^j,}
\end{equation}
here $\mathcal{U}_{\rm{c}}$ is the quantity $\mathcal{U}$ at the center.
Since both GR and its Newtonian counterpart with small $\phi$ and $\x$ are nonlinear, the TOV equations are also nonlinear.
Due to the more involved nonlinearity of the TOV equations, one often solves them by adopting numerical algorithms via a selected $\epsc$ and an input dense matter EOS\,\citep{CaiLi2016,LiF2022} as well as the termination condition:
\begin{equation}\label{def-Radius}
    P(R)=0\leftrightarrow\hP(\widehat{R})=0,
\end{equation}
which defines the NS radius $R$.
The NS mass is given as 
\begin{equation}\label{Mrr}
M_{\rm{NS}}=\widehat{M}_{\rm{NS}}W,~~\mbox{with}~~\widehat{M}_{\rm{NS}}\equiv \widehat{M}(\widehat{R})=\int_0^{\widehat{R}}\d\hr\hr^2\widehat{\varepsilon}(\hr).
\end{equation}

Starting from the scaled TOV equations of (\ref{TOV-ds}), we can show that both $\hP$ and $\heps$ are even under the transformation $\hr\leftrightarrow-\hr$ while $\hM$ is odd\,\citep{CL24-a}.
Therefore,  we can write down the general expansions for $\widehat{\varepsilon}$,  $\widehat{P}$ and $\widehat{M}$ near $\widehat{r}=0$:
\begin{empheq}[box=\fbox]{align}
\heps(\hr)\approx&1+a_2\hr^2+a_4\hr^4+a_6\hr^6+\cdots,\label{ee-heps}\\
\hP(\hr)\approx&\x+b_2\hr^2+b_4\hr^4+b_6\hr^6+\cdots,\label{ee-hP}\\
\hM(\hr)\approx &\frac{1}{3}\hr^3+\frac{1}{5}a_2\hr^5+\frac{1}{7}a_4\hr^7+\frac{1}{9}a_6\hr^9+\cdots,\label{ee-hM}
\end{empheq}
the expansion for $\hM$ follows directly from that for $\heps$.
As a direct consequence, we find that $s^2(\hr)=s^2(-\hr)$, i.e.,  there would be no odd terms in $\hr$ in the expansion of $s^2$ over $\hr$.
The relationships between $\{a_j\}$ and $\{b_j\}$ are determined by the scaled TOV equations of (\ref{TOV-ds}); and the results are\,\citep{CLZ23-a}
\begin{align}
b_2=&-\frac{1}{6}\left(1+3\widehat{P}_{\rm{c}}^2+4\widehat{P}_{\rm{c}}\right),\label{ee-b2}\\
b_4=&\frac{\widehat{P}_{\rm{c}}}{12}\left(1+3\widehat{P}_{\rm{c}}^2+4\widehat{P}_{\rm{c}}\right)
-\frac{a_2}{30}\left(4+9\widehat{P}_{\rm{c}}\right),\label{ee-b4}\\
b_6=&-\frac{1}{216}\left(1+9\widehat{P}_{\rm{c}}^2\right)\left(1+3\widehat{P}_{\rm{c}}^2+4\widehat{P}_{\rm{c}}\right)-\frac{a_2^2}{30}+\left(\frac{2}{15}\widehat{P}_{\rm{c}}^2+\frac{1}{45}\widehat{P}_{\rm{c}}-\frac{1}{54}\right)a_2-\frac{5+12\widehat{P}_{\rm{c}}}{63}a_4,\label{ee-b6}
\end{align}
etc., and all the odd terms of $\{b_j\}$ and $\{a_j\}$ are zero.
The coefficient $a_2$ can be expressed in terms of $b_2$ via the SSS, because
\begin{equation}\label{s2_r_exp}
s^2=\frac{\d\widehat{P}}{\d\widehat{\varepsilon}}=\frac{\d\widehat{P}}{\d\widehat{r}}\cdot\frac{\d\widehat{r}}{\d\widehat{\varepsilon}}=\frac{b_2+2b_4\widehat{r}^2+\cdots}{a_2+2a_4\widehat{r}^2+\cdots}.
\end{equation}
Evaluating it at $\hr=0$ gives $s_{\rm{c}}^2=b_2/a_2$,  or inversely $a_2=b_2/s_{\rm{c}}^2$.
Since $s_{\rm{c}}^2>0$ and $b_2<0$, we find $
a_2<0$, i.e., the energy density is a monotonically decreasing function of $\hr$ near $\hr\approx0$.

According to the definition of NS radius given in Eq.\,(\ref{def-Radius}), we obtain from the truncated equation $\x+b_2\widehat{R}^2\approx0$ that $\widehat{R}\approx(-\x/b_2)^{1/2}=[6\x/(1+3\x^2+4\x)]^{1/2}$ and therefore the radius $R$\,\citep{CLZ23-a}:
\begin{equation}
\boxed{
    R=\widehat{R}Q\approx\left(\frac{3}{2\pi G}\right)^{1/2}\nu_{\rm{c}},~~\mbox{with}~~
   \nu_{\rm{c}}\equiv\frac{1}{\sqrt{\varepsilon_{\rm{c}}}}\left(\frac{\x}{1+3\x^2+4\x}\right)^{1/2}
   .}\label{def-nc}
\end{equation}
Similarly, the NS mass scales as\,\citep{CLZ23-a}
\begin{equation}
\boxed{
    M_{\rm{NS}}\approx\frac{1}{3}\widehat{R}^3\widehat{\varepsilon}_{\rm{c}}W=
    \frac{1}{3}\widehat{R}^3W\approx\left(\frac{6}{\pi G^3}\right)^{1/2}\Gamma_{\rm{c}},~~\mbox{with}~~\Gamma_{\rm{c}}
    \equiv\frac{1}{\sqrt{\varepsilon_{\rm{c}}}}\left(\frac{\x}{1+3\x^2+4\x}\right)^{3/2}
    .}\label{def-Gc}
\end{equation}
Consequently, the NS compactness $\xi$ scales as\,\citep{CL24-b}
\begin{equation}
\boxed{
    \xi\equiv\frac{M_{\rm{NS}}}{R}\approx\frac{2}{G}\frac{\x}{1+3\x^2+4\x}
    = \frac{2\Pi_{\rm{c}}}{G},~~\mbox{with}~~\Pi_{\rm{c}}\equiv 
   \frac{\x}{1+3\x^2+4\x}.}\label{def-Pic}
\end{equation}
For small $\x$ (Newtonian limit), $\xi\approx2\x$. The relation (\ref{def-Pic}) implies that $\x$ is the source and also a measure of NS compactness\,\citep{CL24-b}.
The correlation between $\x$ and $\xi$ was studied and fitted numerically in the form of $\ln\x\approx\sum_iz_i\xi^i$ using varius EOS models\,\citep{Saes22}. Such fitting schemes become eventually effective as enough parameters $z_i$'s are used. However, the real correlation between $\x$ and $\xi$ is somehow lost.
In particular, our correlation tells that $\xi\sim\tau_0+\tau_1\x+\tau_2\x^2+\cdots$ with $\tau_0\approx0$ and $\tau_1\approx2$.

\begin{figure}[ht!]
\centering
\includegraphics[width=8.cm]{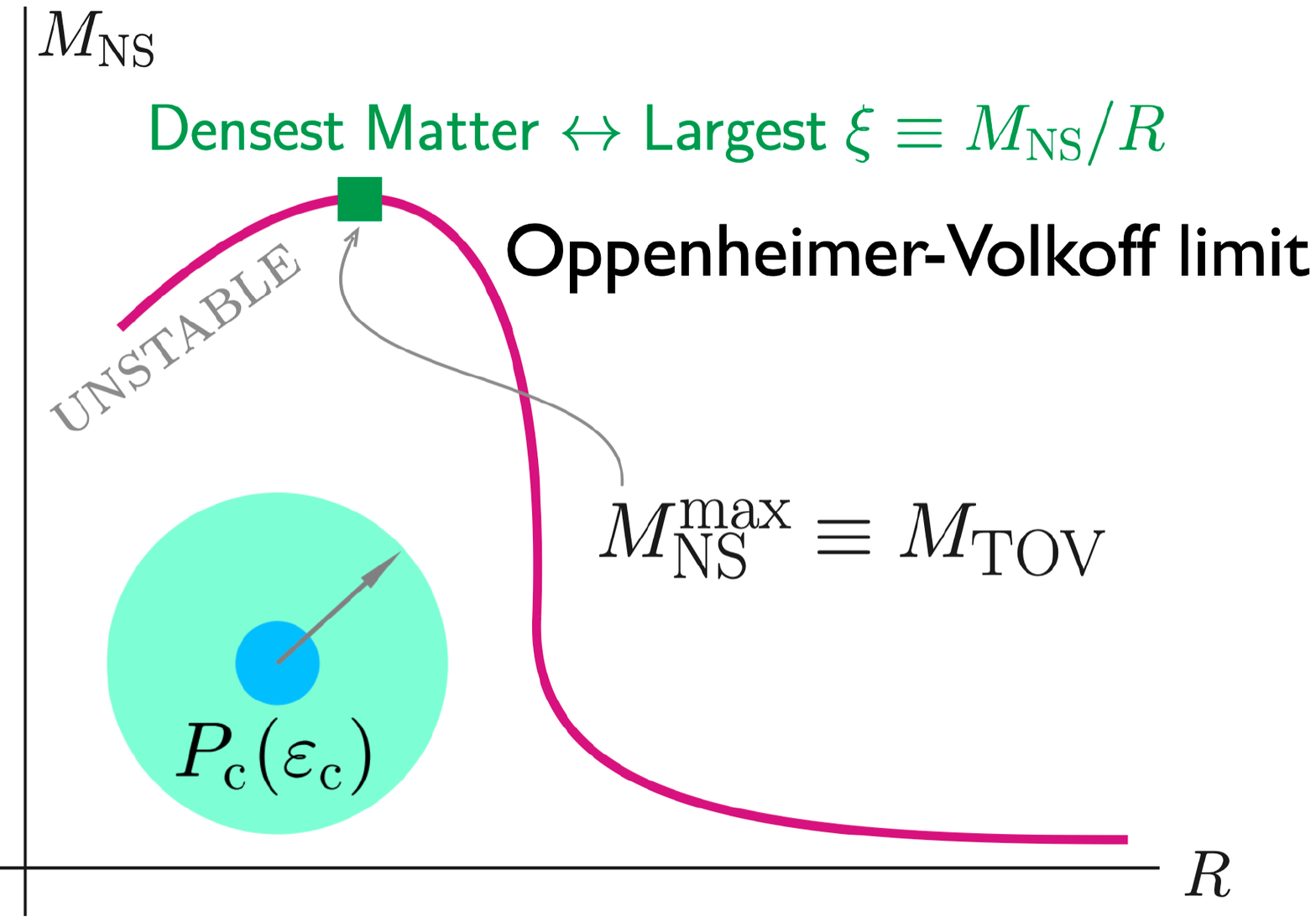}
\caption{\small (Color Online).  An illustration of the TOV configuration on a typical mass-radius sequence. The cores of NSs at the TOV configuration contain the densest visible matter existing in our Universe, the compactness $\xi$ for such NSs is the largest among all stable NSs.}\label{fig_MRTOV}
\end{figure}

The maximum-mass configuration (or the TOV configuration) along the NS M-R curve is a special point.
Consider a typical NS M-R curve near the TOV configuration from right to left, the radius $R$ (mass $M_{\rm{NS}}$) eventually decreases (increases), the compactness $\xi=M_{\rm{NS}}/R$ correspondingly increases and reaches its maximum value at the TOV configuration.
When going to the left along the M-R curve even further, the stars becomes unstable and then may collapse into black holes (BHs).
So the NS at the TOV configuration is denser than its surroundings and the cores of such NSs contain the stable densest visible matter existing in the Universe.
The TOV configuration is indicated on a typical M-R sequence in FIG.\,\ref{fig_MRTOV}. Mathematically, the TOV configuration is described as,
\begin{equation}\label{MTOV-cr}
\left.
\frac{\d M_{\rm{NS}}}{\d\varepsilon_{\rm{c}}}\right|_{M_{\rm{NS}}=M_{\rm{NS}}^{\max}=M_{\rm{TOV}}}
=0.
\end{equation}
Using the NS mass scaling of Eq.\,(\ref{def-Gc}), we obtain 
\begin{equation}
    \frac{\d M_{\rm{NS}}}{\d\varepsilon_{\rm{c}}}=\frac{1}{2}\frac{M_{\rm{NS}}}{\varepsilon_{\rm{c}}}\left[3\left(\frac{s_{\rm{c}}^2}{\x}-1\right)\frac{1-3\x^2}{1+3\x^2+4\x}-1\right],~~\mbox{where}~~s_{\rm{c}}^2\equiv\frac{\d P_{\rm{c}}}{\d\varepsilon_{\rm{c}}}.
\end{equation}
Inversely, we obtain the expression for the central SSS\,\citep{CLZ23-b,CL24-a},
\begin{equation}\label{sc2-GG}
\boxed{
\mbox{for stable NSs along M-R curve:}~~
s_{\rm{c}}^2=\x\left(1+\frac{1+\Psi}{3}\frac{1+3\x^2+4\x}{1-3\x^2}\right),}
\end{equation}
where \begin{equation}\label{def-Psi}
\Psi
=2\frac{\d\ln M_{\rm{NS}}}{\d\ln\varepsilon_{\rm{c}}}\geq0.
\end{equation}
We see that the SSS really has the form of Eq.\,(\ref{def-s2fphi}).
For NSs at the TOV configuration, we have
\begin{equation}\label{sc2-TOV}
\boxed{
\mbox{for NSs at the TOV configuration:}~~
s^2_{\rm{c}}=\x\left(1+\frac{1}{3}\frac{1+3\x^2+4\x}{1-3\x^2}\right).}
\end{equation}
since now $\Psi=0$.
Using the $s_{\rm{c}}^2$ of Eq.\,(\ref{sc2-TOV}) for NSs at the TOV configuration, we can calculate the derivative of NS radius $R$ with respective to $\varepsilon_{\rm{c}}$ around the TOV point, that is\,\citep{CLZ23-a}
\begin{equation}
\frac{\d R}{\d\varepsilon_{\rm{c}}}\sim
\frac{\d}{\d\varepsilon_{\rm{c}}}\left(\frac{\widehat{R}}{\sqrt{\varepsilon_{\rm{c}}}}\right)_{R_{\max}\leftrightarrow M_{\rm{NS}}^{\max}}
=\left(\frac{s_{\rm{c}}^2}{\x}-1\right)\frac{1-3\x^2}{1+3\x^2+4\x}-1=-\frac{2}{3},
\end{equation}
i.e., as $\varepsilon_{\rm{c}}$ increases, the radius $R$ decreases (self-gravitating property), as expected.
On the other hand, for stable NSs along the M-R curve with a nonzero $\Psi$, we have $\d R/\d\varepsilon_{\rm{c}}\sim(\Psi-2)/3$; this means if $\Psi$ is around 2, the dependence of the radius on $\varepsilon_{\rm{c}}$ would be weak.
\begin{figure}[ht!]
\centering
\includegraphics[height=9.cm]{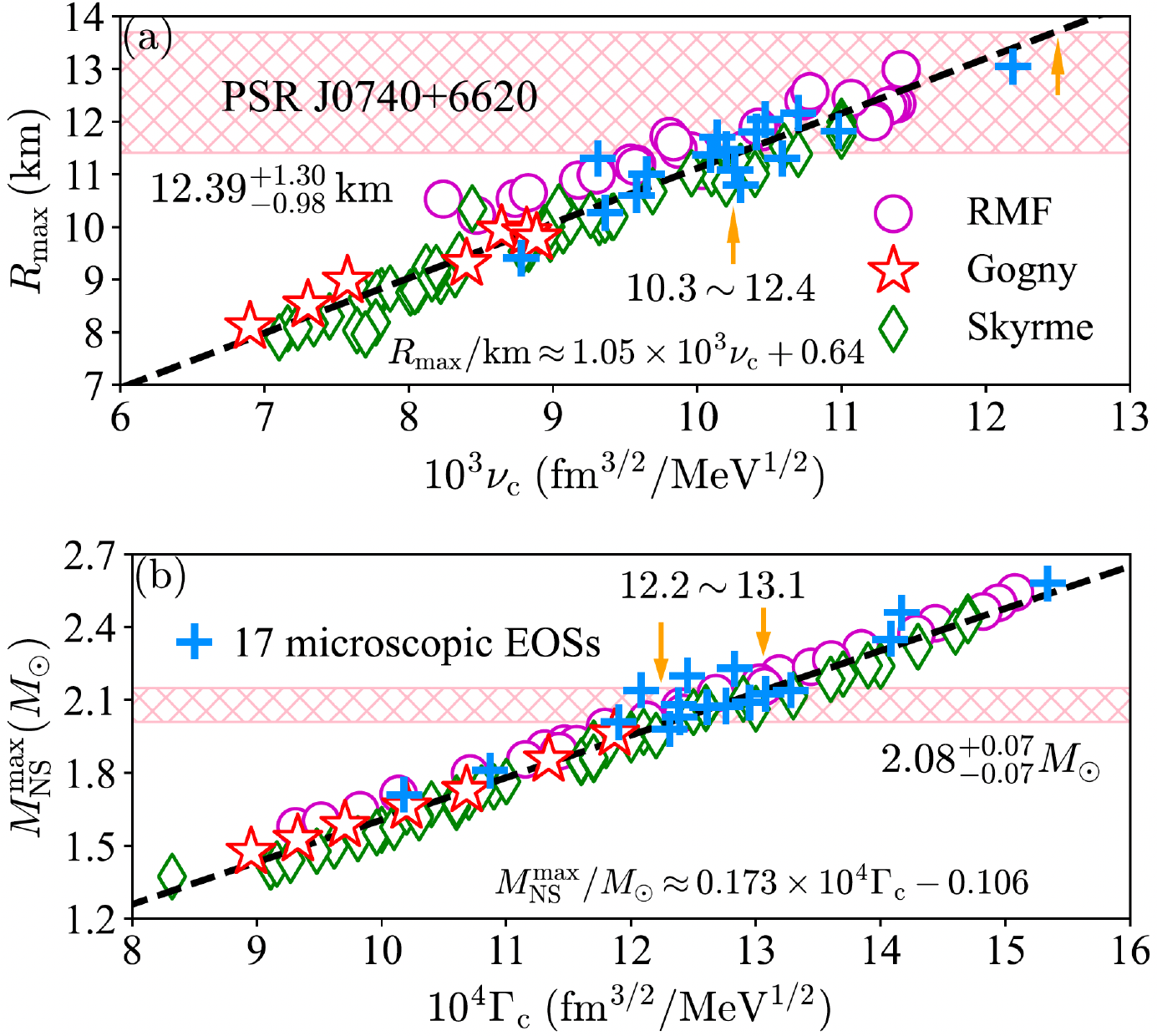}
\caption{\small (Color Online). 
Panel (a): the $R_{\max}$-$\nu_{\rm{c}}$ correlation
using 104 EOS samples (colored symbols), see Ref.\,\cite{CLZ23-a} for more detailed descriptions on these EOSs,  the constraints on the mass\,\citep{Fon21} and radius\,\citep{Riley21} of PSR J0740+6620 are shown by the pink hatched bands. Panel (b): similar as the left panel but fot $M_{\rm{NS}}^{\max}$-$\Gamma_{\rm{c}}$.
The orange arrows and captions nearby in each panel indicate the $\nu_{\rm{c}}$ and $\Gamma_{\rm{c}}$ defined in Eq.\,(\ref{def-nc}) and Eq.\,(\ref{def-Gc}), respectively.
Figures taken from Ref.\,\cite{CLZ23-a}.
}\label{fig_MmaxS}
\end{figure}

For verification, the scaling $R_{\max}$-$\nu_{\rm{c}}$ (panel (a)) of Eq.\,(\ref{def-nc}) and the scaling $M_{\rm{NS}}^{\max}$-$\Gamma_{\rm{c}}$ (panel (b)) of Eq.\,(\ref{def-Gc}) are shown in FIG.\,\ref{fig_MmaxS} by using 87 phenomenological and 17 extra microscopic NS EOSs with and/or without considering hadron-quark phase transitions and hyperons by solving numerically the original TOV equations, see Ref.\,\cite{CLZ23-a} for more details on these EOS samples.
The observed strong linear correlations demonstrate vividly that the $R_{\max}$-$\nu_{\rm{c}}$ and $M_{\rm{NS}}^{\max}$-$\Gamma_{\rm{c}}$ scalings are nearly universal. While it is presently unclear where the mass threshold for massive NSs to collapse into BHs is located, 
the TOV configuration is the closest to it theoretically. It is also well known that certain properties of BHs are universal and only depend on quantities like mass, charge and angular momentum. One thus expects the NS mass and radius scalings near the TOV configuration to be more EOS-independent compared to light NSs. It is also particularly interesting to notice that EOSs allowing phase transitions and/or hyperon formations predict consistently the same scalings. 

By performing linear fits of the results obtained from the EOS samples, the quantitative scaling relations are\,\citep{CLZ23-a,CLZ23-b,CL24-a}
\begin{align}
R_{\rm{max}}/\rm{km}
\approx& 1050_{-30}^{+30}\times\left(\frac{\nu_{\rm{c}}}{\rm{fm}^{3/2}/\rm{MeV}^{1/2}}\right)+0.64_{-0.25}^{+0.25},\label{Rmax-n}\\
M_{\rm{NS}}^{\rm{max}}/M_{\odot} 
\approx& 1730_{-30}^{+30} \times \left(\frac{\Gamma_{\rm{c}}}{\rm{fm}^{3/2}/\rm{MeV}^{1/2}}\right)-0.106_{-0.035}^{+0.035},\label{Mmax-G}
\end{align} with their Pearson's coefficients about 0.958 and 0.986, respectively, here $\nu_{\rm{c}}$ and $\Gamma_{\rm{c}}$ are measured in $\rm{fm}^{3/2}/\rm{MeV}^{1/2}$.
In addition, the standard errors for the radius and mass fittings are about 0.031 and 0.003 for these EOS samples.
In FIG.\,\ref{fig_MmaxS}, the condition $M^{\max}_{\rm{NS}}\gtrsim1.2M_{\odot}$ used is necessary to mitigate influences of uncertainties in modeling the crust EOS\,\citep{BPS71,Iida1997,XuJ} for low-mass NSs. For the heavier NSs studied here, it is reassuring to see that although the above 104 EOSs predicted quite different crust properties, they all fall closely around the same scaling lines consistently, especially for the $M_{\rm{NS}}^{\max}$-$\Gamma_{\rm{c}}$ relation.

\section{Gravitational Upper Bound on \texorpdfstring{$\x\equiv\phi_{\rm{c}}=P_{\rm{c}}/\varepsilon_{\rm{c}}$}{\texttwoinferior}, its Generalizations and the Impact on Supradense NS Matter EOS}
\label{SEC_3}

Based on Eq.\,(\ref{sc2-TOV}) and the Principle of Causality of SR, we obtain immediately\,\citep{CLZ23-a}
\begin{equation}\label{Xupper}
\boxed{
s_{\rm{c}}^2\leq1\leftrightarrow\x=\widehat{P}_{\rm{c}}\lesssim0.374\equiv\x_+^{\rm{GR}}.}
\end{equation}
Although the causality condition requires apparently $\widehat{P}_{\rm{c}}\leq1$, the supradense nature of core NS matter indicated by the nonlinear dependence of $s_{\rm{c}}^2$ on $\widehat{P}_{\rm{c}}$ essentially renders it to be much smaller.

A small $\x<1$ was in fact studied/indicated earlier in the literature\,\citep{Koranda1997,Saes22}.
For example, in Ref.\,\cite{Koranda1997}, the minimum-period EOS of the form $P(\varepsilon)=0$ for $\varepsilon<\varepsilon_{\rm{f}}$ and $P(\varepsilon)=\varepsilon-\varepsilon_{\rm{f}}$ for $\varepsilon\geq\varepsilon_{\rm{f}}$ 
 was adopted; here $\varepsilon_{\rm{f}}$ is a free parameter of the model. Such EOS is simplified and unrealistic in the sense: (1) both the parameter $\varepsilon_{\rm{f}}\approx2.156\times10^{15}\,\rm{g}/\rm{cm}^3\approx8.1\varepsilon_0$ and the central energy density $\varepsilon_{\rm{c}}\approx4.778\times10^{15}\rm{g}/\rm{cm}^3\approx17.9\varepsilon_0$ are unrealistically large for a 1.442$M_{\odot}$ NS\,\citep{Koranda1997}; the consequent ratio $\x$ in this model is $\x=1-\varepsilon_{\rm{f}}/\varepsilon_{\rm{c}}\approx0.55$; (2) the central SSS of 1 of such model is basically inconsistent with Eq.\,(\ref{sc2-TOV}).
 Actually, only with $\x=1-\varepsilon_{\rm{f}}/\varepsilon_{\rm{c}}\approx0.374$ or $\varepsilon_{\rm{f}}/\varepsilon_{\rm{c}}\approx0.626$ one can make this EOS model consistent with Eq.\,(\ref{sc2-TOV}), i.e., the parameter space for $\varepsilon_{\rm{f}}$ is limited; however a vanishing pressure up to $\varepsilon_{\rm{f}}/\varepsilon_{\rm{c}}\approx0.626$ is fundamentally unsatisfactory. 
 Therefore, $\x\approx0.55$ is only qualitatively meaningful.

The bound (\ref{Xupper}) is obtained under the specific condition that it gives the upper limit for $\phi=P/\varepsilon$ at the center of NSs at TOV configurations.
In order to bound a general $\phi=P/\varepsilon=\widehat{P}/\widehat{\varepsilon}$, we need to take three generalizations of $\x\lesssim0.374$ obtained from Eq.\,(\ref{Xupper}) by asking\,\citep{CLZ23-b},
\begin{itemize}
\item[(a)] How does $\phi=\widehat{P}/\widehat{\varepsilon}$ behave at a finite $\widehat{r}$ for the maximum-mass configuration $M_{\rm{NS}}^{\max}$?
\item[(b)] How does the limit $\x\lesssim0.374$ modify when considering stable NSs on the M-R curve away from the TOV configuration?
\item[(c)] By combining (a) and (b), how does $\phi$ behave for stable NSs at finite distances $
\widehat{r}$ away from their centers?
\end{itemize}
For the first question,  since the pressure $\widehat{P}$ and $\widehat{\varepsilon}$ are both decreasing functions of $\widehat{r}$, i.e., $
\widehat{P}\approx \widehat{P}_{\rm{c}}+b_2\widehat{r}^2<\widehat{P}_{\rm{c}}$ and $
\widehat{\varepsilon}\approx1+s_{\rm{c}}^{-2}b_2\widehat{r}^2<1$ (notice $\widehat{\varepsilon}_{\rm{c}}=1$ and $a_2=b_2/s_{\rm{c}}^2$), 
we obtain by taking their ratio:
\begin{align}
\phi=&{P}/{\varepsilon}=
{\widehat{P}}/{\widehat{\varepsilon}}\approx{\widehat{P}_{\rm{c}}}/{\widehat{\varepsilon}_{\rm{c}}}+\left(1-\frac{\widehat{P}_{\rm{c}}}{s_{\rm{c}}^2}\right)b_2\widehat{r}^2
={\widehat{P}_{\rm{c}}}+\left(1-\frac{\widehat{P}_{\rm{c}}}{s_{\rm{c}}^2}\right)b_2\widehat{r}^2
\approx\widehat{P}_{\rm{c}}-\left(\frac{1+7\widehat{P}_{\rm{c}}}{24}\right)\widehat{r}^2
<\widehat{P}_{\rm{c}}.\label{pk-3}
\end{align}
Generally, $1-\widehat{P}_{\rm{c}}/s_{\rm{c}}^2>0$, the small-$\widehat{P}_{\rm{c}}$ expansions of $s_{\rm{c}}^2$ of Eq.\,(\ref{sc2-TOV}) and $b_2$ of Eq.\,(\ref{ee-b2}) are used in the last step.
This means that not only $\widehat{P}$ and $\widehat{\varepsilon}$ decrease for finite $\widehat{r}$,  but also does their ratio $\widehat{P}/\widehat{\varepsilon}$.
Therefore for NSs at the TOV configuration of the M-R curves, we have $\phi=\widehat{P}/\widehat{\varepsilon}\leq\widehat{P}_{\rm{c}}\lesssim0.374$.
Considering the second question and for stable NSs on the M-R curve, one has $\Psi>0$ (of Eq.\,(\ref{def-Psi})) and Eq.\,(\ref{sc2-GG}) induces an even smaller upper bound for $\x$ compared with 0.374.
Furthermore, for the last question (c),  the inequality (\ref{pk-3}) still holds and is slightly modified for small $\widehat{P}_{\rm{c}}$ as,
\begin{equation}\label{pk-3a}
\phi= \widehat{P}/\widehat{\varepsilon}\approx\widehat{P}_{\rm{c}}
-\frac{1}{24}\frac{1+\Psi}{(1+\Psi/4)^2}\left[1+7\widehat{P}_{\rm{c}}+\Psi\left(\widehat{P}_{\rm{c}}+\frac{1}{4}\right)\right]\hr^2
<\widehat{P}_{\rm{c}},
 \end{equation}
which implies $\phi=\widehat{P}/\widehat{\varepsilon}$ for $\Psi\neq0$ also decreases with $\hr$.

Combining the above three aspects, we find 
\begin{equation}\label{Xupper-GEN}
\boxed{
\mbox{for stable NSs along M-R curve near/at the centers:}~~
\phi=P/\varepsilon=\widehat{P}/\widehat{\varepsilon}\leq \x\lesssim0.374.}
\end{equation}
Nevertheless, the validity of this conclusion is limited to small $\widehat{r}$ due to the perturbtive nature of the expansions of $\widehat{P}(\widehat{r})$ and $\widehat{\varepsilon}(\widehat{r})$. Whether $\phi=P/\varepsilon$ could exceed such upper limit at even larger distances away from the centers depends on the joint analysis of $s^2$ and $P/\varepsilon$,  e.g., by including more higher order contributions of the expansions\,\citep{CLZ23-b}.
The upper bound $P/\varepsilon\lesssim0.374$ (at least near the NS centers) is an intrinsic property of the TOV equations, which embody the strong-field aspects of gravity in GR, especially the strong self-gravitating nature.
In this sense, there is no guarantee {\it a prior} that this bound is consistent with all microscopic nuclear EOSs (either relativistic or non-relativistic). This is mainly because the latter were conventionally constructed without considering the strong-field ingredients of gravity.
The robustness of such upper bound for $\phi=P/\varepsilon$ can be checked only by observable astrophysical quantities/processes involving strong-field aspects of gravity such as NS M-R data, NS-NS mergers and/or NS-BH mergers\,\citep{BS2010,Shibata2015,Baiotti2017,Kyutoku2021}. As mentioned earlier, in the NS matter-gravity inseparable system, it is the total action that determines the matter state and NS structure. Thus, to our best knowledge, there is no physics requirement that the EOS of supradense matter created in vacuum from high-energy heavy-ion collisions or other laboratory experiments where effects of gravity can be neglected to be the same as that in NSs as nuclear matter in the two situations are in very different environments. Nevertheless, ramifications of the above findings and logical arguments should be further investigated.

\begin{figure}[ht!]
\centering
\includegraphics[width=15.cm]{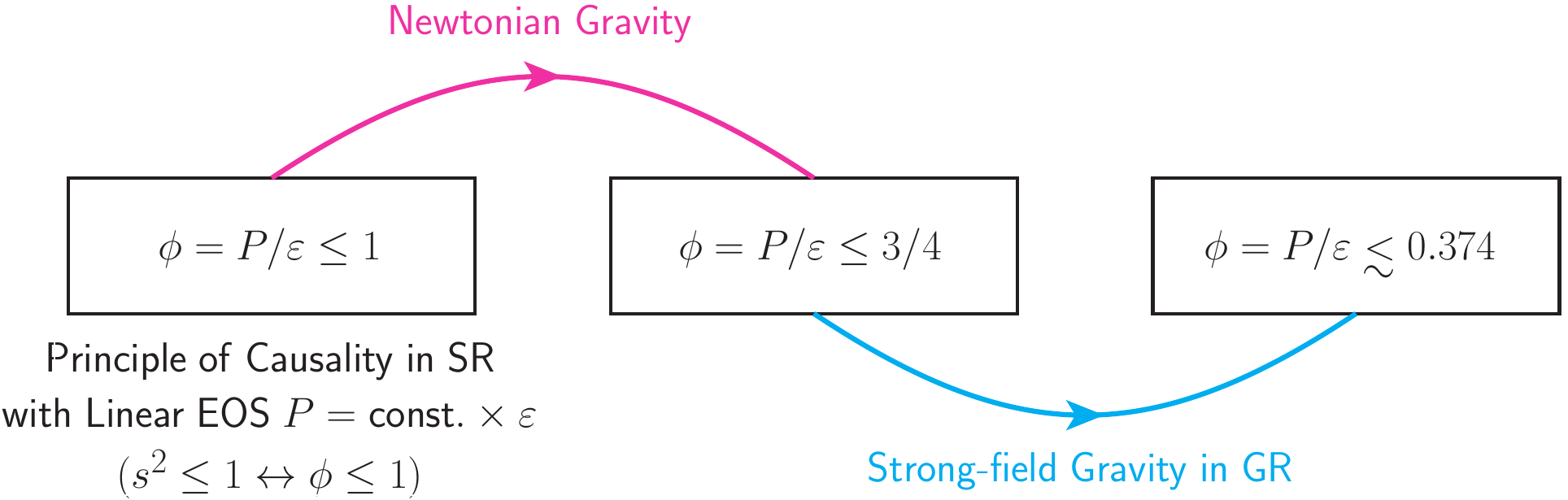}
\caption{\small (Color Online). An illustration of gravitational effects on supradense matter EOS in NSs: The nonlinearity of Newtonian gravity reduces the upper bound for $\phi$ from 1 (obtained by requiring $s^2\leq1$ in SR via a linear EOS of the form $P=\rm{const.}\times\varepsilon$ for supradense matter in vacuum) to 3/4=0.75 and the even stronger nonlinearity of the gravity in GR further refines it to be about 0.374.
}\label{fig_phi-kkk}
\end{figure}

Next, we consider the Newtonian limit where $\phi$ and $\x$ are small, then we can neglect $3\x^2+4\x$ in the coefficient $b_2$, consequently $b_2=-1/6$ is obtained\,\citep{Chan10-a}.
In such case, we shall obtain from Eq.\,(\ref{sc2-TOV}):
\begin{equation}
\boxed{
\mbox{Newtonian limit:}~~
    s_{\rm{c}}^2\approx{4\x}/{3},}
\end{equation}
and the principle of causality requires $\x\leq3/4=0.75\equiv\x_+^{\rm{N}}$. The latter can be applied to nuclear matter created in laboratory experiments where effects of gravity can be neglected. Turning on gravity in NSs, we see that the nonlinearity of Newtonian gravity has already reduced the upper bound for $\phi$ from 1 obtained by requiring $s^2\leq1$ in SR via a linear EOS of the form $P=\rm{const.}\times \varepsilon$ to 3/4, the even stronger nonlinearity of the gravity in GR reduces it further. These effects of gravity on $\phi$ are illustrated in FIG.\,\ref{fig_phi-kkk}. 
It is seen that the strong-field gravity in GR brings a relative reduction on the upper bound for $\phi$ by about 100\%.
Though the $\phi$ or $\x$ in Newtonian gravity is generally smaller, the upper bound for $\phi$ or $\x$ is however larger than its GR counterpart. The index $s_{\rm{c}}^2/\x$ being greater than 1 in both Newtonian gravity and in GR imply that the central EOS in NSs once considering the gravity effect could not be linear or conformal.

We emphasize that all of the analyses above based on SR and GR are general from analyzing perturbatively analytical solutions of the scaled TOV equations without using any specific nuclear EOS. Because the TOV equations are results of a hydrodynamical equilibrium of NS matter in the environment of a strong-field gravity from extremizing the total action of the matter-gravity system, features revealed above from SR and GR inherent in the TOV equations must be matched by the nuclear EOS. This requirement can then put strong constraints on the latter. In particular, the upper bound for $\phi$ as $\phi\lesssim\x_+^{\rm{GR}}\approx0.374$ of Eq.\,(\ref{Xupper}) enables us to limit the density dependence of nuclear EOS relevant for NS modelings. 

\begin{figure}[ht!]
\centering
\includegraphics[width=10.cm]{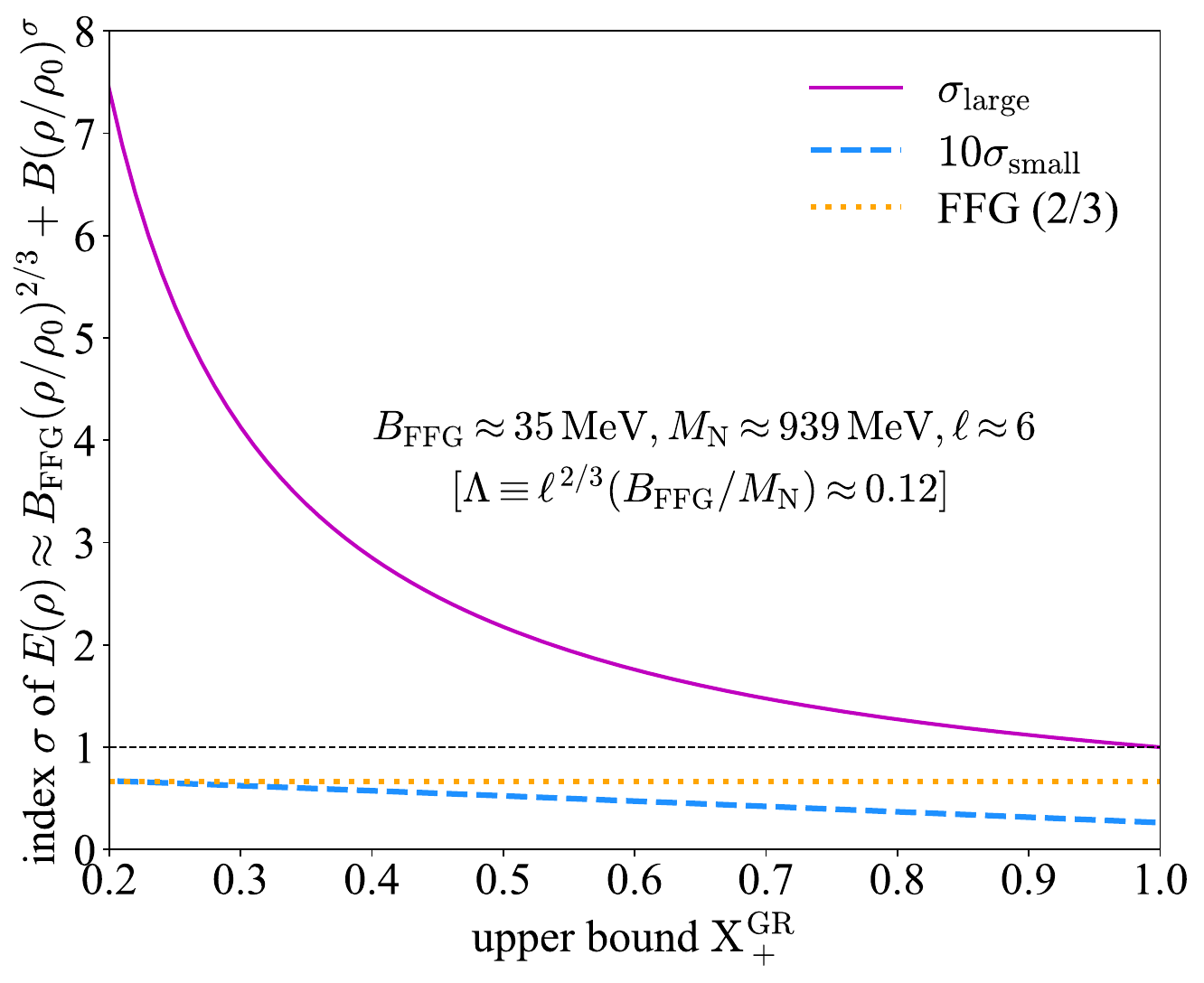}
\caption{\small (Color Online). Gravitational impact on the EOS of supradense matter and the underlying strong interaction in NSs: the general $\x_+^{\rm{GR}}$-dependence of $\sigma_{\rm{large}}$ and $\sigma_{\rm{small}}$ of Eq.\,(\ref{dq-2}), based on the nuclear EOS model of Eq.\,(\ref{Edexp-1}); here $B_{\rm{FFG}}\approx35\,\rm{MeV}$, $M_{\rm{N}}\approx939\,\rm{MeV}$ and $\ell=\rho/\rho_0\approx6$.
}\label{fig_sigmaX}
\end{figure}

In the following, we provide an example illustrating how the strong-field gravity can restrict the behavior of superdense matter in NSs.  For simplicity, we assume that the energy per baryon takes the following form
\begin{equation}\label{Edexp-1}
    E(\rho)=B_{\rm{FFG}}\left(\frac{\rho}{\rho_0}\right)^{2/3}+B\left(\frac{\rho}{\rho_0}\right)^{\sigma},
\end{equation}
where the first term is the kinetic energy of a FFG of neutrons in NSs with $B_{\rm{FFG}}\approx 35$ MeV being its known value at $\rho_0$ and the second term is the contribution from interactions described with the parameters $B$ and $\sigma$. The pressure and the energy density are obtained from $P(\rho)=\rho^2\d E/\d\rho$ and $\varepsilon(\rho)=[E(\rho)+M_{\rm{N}}]\rho$, respectively.
The ratio $\phi=P/\varepsilon$ and the SSS $s^2=\d P/\d\varepsilon$ could be obtained correspondingly.
Denote the reduced density $\rho/\rho_0$ where $s^2\to1$ and $\phi\to\x\to\x_+^{\rm{GR}}$ as $\ell$ (e.g., $\ell\lesssim8$ for realistic NSs), the following constraining equation for $\sigma$ is obtained:
\begin{equation}\label{dq-1}
    \sigma\left(\x_+^{\rm{GR}}\sigma-1\right)+\frac{\ell^{2/3}}{3}\left(\frac{B_{\rm{FFG}}}{M_{\rm{N}}}\right)
    \left(\sigma-\frac{2}{3}\right)\left[
    \left(3\sigma+2\right)\x_+^{\rm{GR}}-2\sigma-3
    \right]=0.
\end{equation}
Thus, $\x_+^{\rm{GR}}$ effectively restricts the index $\sigma$ characterizing the stiffness of nuclear EOS.
There are two solutions of Eq.\,(\ref{dq-1}) with one being greater than 1 (denoted as $\sigma_{\rm{large}}$) and the other smaller than 1 (denoted as $\sigma_{\rm{small}}$). They can be explicitly written as
\begin{align}
  \sigma=\frac{1}{2}\left(\x_+^{\rm{GR}}
  +\Lambda\left(\x_+^{\rm{GR}}-\frac{2}{3}\right)\right)^{-1}
  \left\{1+\frac{5}{9}\Lambda\pm\sqrt{1+\frac{16\Lambda}{9}
  \left[\left(\x_+^{\rm{GR},2}-\frac{3\x_+^{\rm{GR}}}{2}+\frac{5}{8}\right)+\Lambda\left(\x_+^{\rm{GR}}-\frac{13}{12}\right)^2\right]
  }\right\},\label{dq-2}
\end{align}
where 
\begin{equation}
    \Lambda\equiv\ell^{2/3}\left(\frac{B_{\rm{FFG}}}{M_{\rm{N}}}\right)\ll1.
\end{equation}
The expression for the coefficient $B$ is
\begin{equation}\label{dq-3}
    B=\left(\frac{1+5\Lambda/9}{\sigma^2-1}\frac{1}{\ell^{\sigma}}\right)M_{\rm{N}},
\end{equation}
which depends on $\x_+^{\rm{GR}}$ through $\sigma$.
As a numerical example, using $M_{\rm{N}}\approx939\,\rm{MeV}$, $B_{\rm{FFG}}\approx35\,\rm{MeV}$ and $\ell\approx6$ leads to $\sigma_{\rm{large}}\approx3.1$ and $B_{\rm{large}}\approx0.45\,\rm{MeV}$ or $\sigma_{\rm{small}}\approx0.06$ and $B_{\rm{small}}\approx-906\,\rm{MeV}$ (this second solution is unphysical since $B>0$ is necessarily required to make $E(\rho)>0$ at NS densities).
If one takes artificially $\x_+^{\rm{GR}}=1$, then the two solutions (\ref{dq-2}) approach 
\begin{equation}
    \sigma_{\rm{small}}\to\frac{2}{3}\frac{1}{1+3/\Lambda}=\frac{2}{3}\left(1+\frac{3}{\ell^{2/3}}\left(\frac{M_{\rm{N}}}{B_\rm{FFG}}\right)\right)^{-1}\ll1, \mbox{~~and~~}\sigma_{\rm{large}}\to1\mbox{~~from above}.
\end{equation}
Now, neither solution is physical since $B_{\rm{small}}<0$ for $\sigma_{\rm{small}}$ while $B_{\rm{large}}\to+\infty$ for $\sigma_{\rm{large}}\to1$ from above, according to Eq.\,(\ref{dq-3}).
The general $\x_+^{\rm{GR}}$-dependence of $\sigma_{\rm{large}}$ and $\sigma_{\rm{small}}$ of Eq.\,(\ref{dq-2}) is shown in FIG.\,\ref{fig_sigmaX}. It is seen that only as $\x_+^{\rm{GR}}\to1$ the EOS approaches a linear form $E(\rho)\approx B\rho/\rho_0\sim\rho$ (so $P\approx B\rho^2/\rho_0$ and $\varepsilon\approx B\rho^2/\rho_0+M_{\rm{N}}\rho$) at large densities (magenta line), being consistent with our general analyses and expectation.

Since the parameterization (\ref{Edexp-1}) is over-simplified, for general cases more density-dependent terms should be included, i.e., $B(\rho/\rho_0)^{\sigma}\to\sum_{j=1}^JB_j(\rho/\rho_0)^{\sigma_j}$. We may then obtain two related equations from $\phi\to\x\to\x_+^{\rm{GR}}$ and $s^2\to1$ as (for either $\x_+^{\rm{GR}}=1$ or $\x_+^{\rm{GR}}\neq1$):
\begin{empheq}[box=\fbox]{align}
    \sum_{j=1}^{J}\left(\frac{B_j}{M_{\rm{N}}}\right)\left(\sigma_j-\x_+^{\rm{GR}}\right)\ell^{\sigma_j}+\ell^{2/3}\left(\frac{B_{\rm{FFG}}}{M_{\rm{N}}}\right)\left(\frac{2}{3}-\x_+^{\rm{GR}}\right)-\x_+^{\rm{GR}}=0,\\
    \sum_{j=1}^J\left(\frac{B_j}{M_{\rm{N}}}\right)\left(1-\sigma_j^2\right)\ell^{1/3+\sigma_j}-\ell^{1/3}\left(1+\frac{5}{9}\ell^{2/3}\left(\frac{B_{\rm{FFG}}}{M_{\rm{N}}}\right)\right)=0.
\end{empheq}
These constraints for $B_j$ and $\sigma_j$ should be taken appropriately into account when writing an effective NS EOS based on density expansions.
For example, when extending Eq.\,(\ref{Edexp-1}) to be $E(\rho)=B_{\rm{FFG}}(\rho/\rho_0)^{2/3}+B_1(\rho/\rho_0)^{\sigma_1}+B_2(\rho/\rho_0)^{\sigma_2}$ under two conditions $E(\rho_0,\delta)\approx E_0(\rho_0)+E_{\rm{sym}}(\rho_0)\delta^2\approx15\,\rm{MeV}$ for pure neutron matter with $\delta=1$ and $P(\rho_0)\approx3\,\rm{MeV}/\rm{fm}^3$, using $\ell\approx6$ together with $\x_+^{\rm{GR}}\approx0.374$, we may obtain $\sigma_1\approx0.3$ and $\sigma_2\approx3.0$ (as well as $B_1\approx-20.5\,\rm{MeV}$ and $B_2\approx0.5\,\rm{MeV}$), respectively. From this example, one can see quantitatively that the gravitational bound naturally leads to a constraint on the nuclear EOS and the underlying interactions in NSs.

\section{Gravitational Lower Bound on Trace Anomaly \texorpdfstring{$\Delta$}{\texttwoinferior} in Supradense NS Matter}\label{SEC_DELTA}

\begin{figure}[ht!]
\centering
\includegraphics[width=10.5cm]{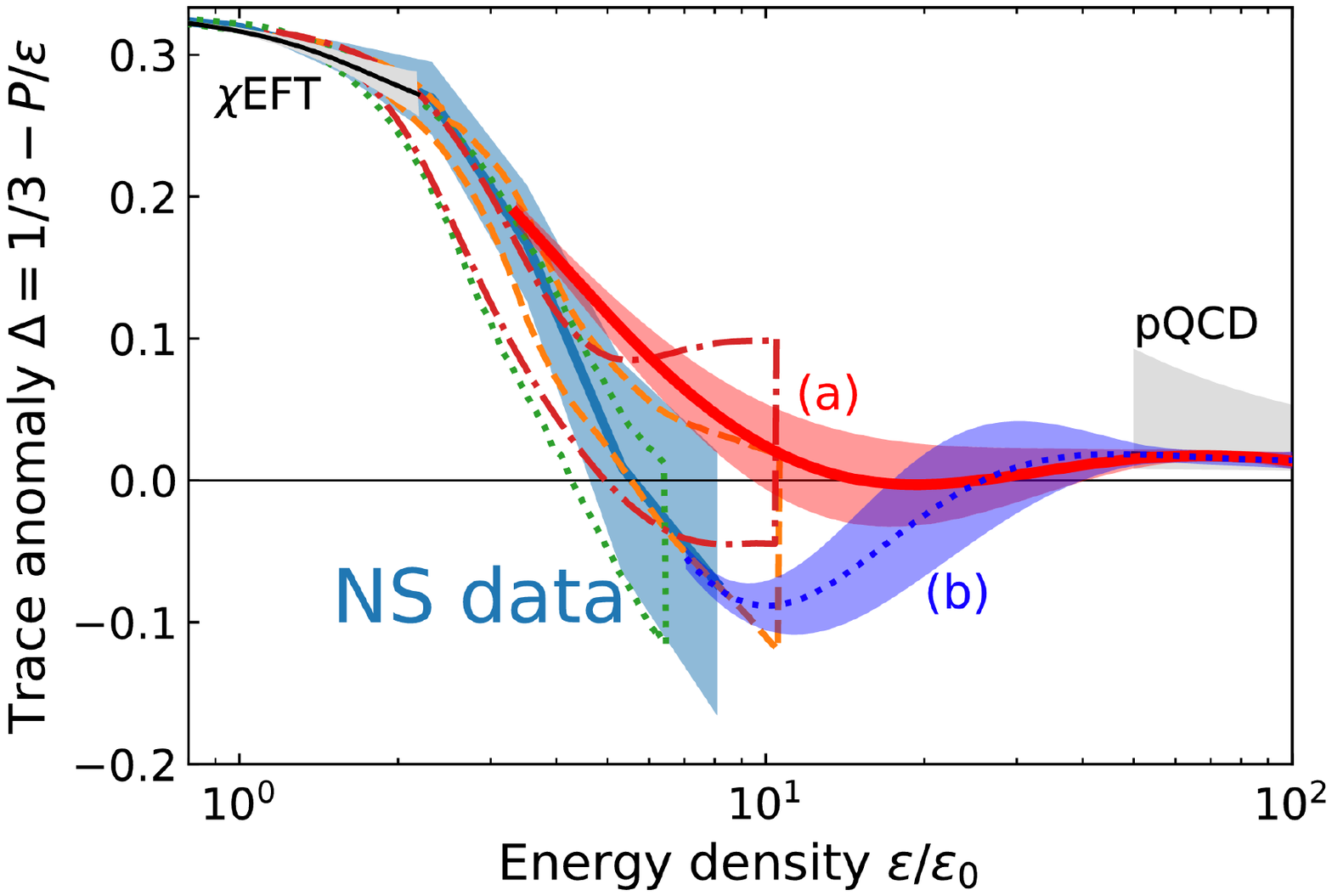}
\caption{\small (Color Online).  Trace anomaly $\Delta$ as a function of energy density $\varepsilon/\varepsilon_0$, here the $\Delta$ in NSs tends to be negative although the pQCD prediction on it approaches zero, $\varepsilon_0\approx150\,\rm{MeV}/\rm{fm}^3$ is the energy density at nuclear saturation density.
Figure taken from Ref.\,\cite{Fuji22}.
}\label{fig_Fuji22Delta}
\end{figure}

After the above general demonstration on the gravitational upper limit for $\phi$ near NS centers given by (\ref{Xupper}) or (\ref{Xupper-GEN}), we equivalently obtain a lower limit on the dimensionless trace anomaly $\Delta=1/3-\phi$ as
\begin{equation}\label{GRDelta}
\boxed{
\Delta\geq \Delta_{\rm{GR}}\approx-0.04.}
\end{equation}
It is very interesting to notice that such GR bound on $\Delta$ is very close to the one predicted by perturbative QCD (pQCD) at extremely high densities owning to the realization of approximate conformal symmetry of quark matter\,\citep{Bjorken83,Fuji22}, as shown in FIG.\,\ref{fig_Fuji22Delta} using certain NS modelings.
A possible negative $\Delta$ in NSs was first pointed out in Ref.\,\cite{Fuji22}, since then several studies have been made on this issue. In the following, we summarize the main findings of these studies by others and compare with what we found above when it is possible. 

\begin{figure}[ht!]
\centering
\hspace*{0.2cm}
\includegraphics[height=9.5cm]{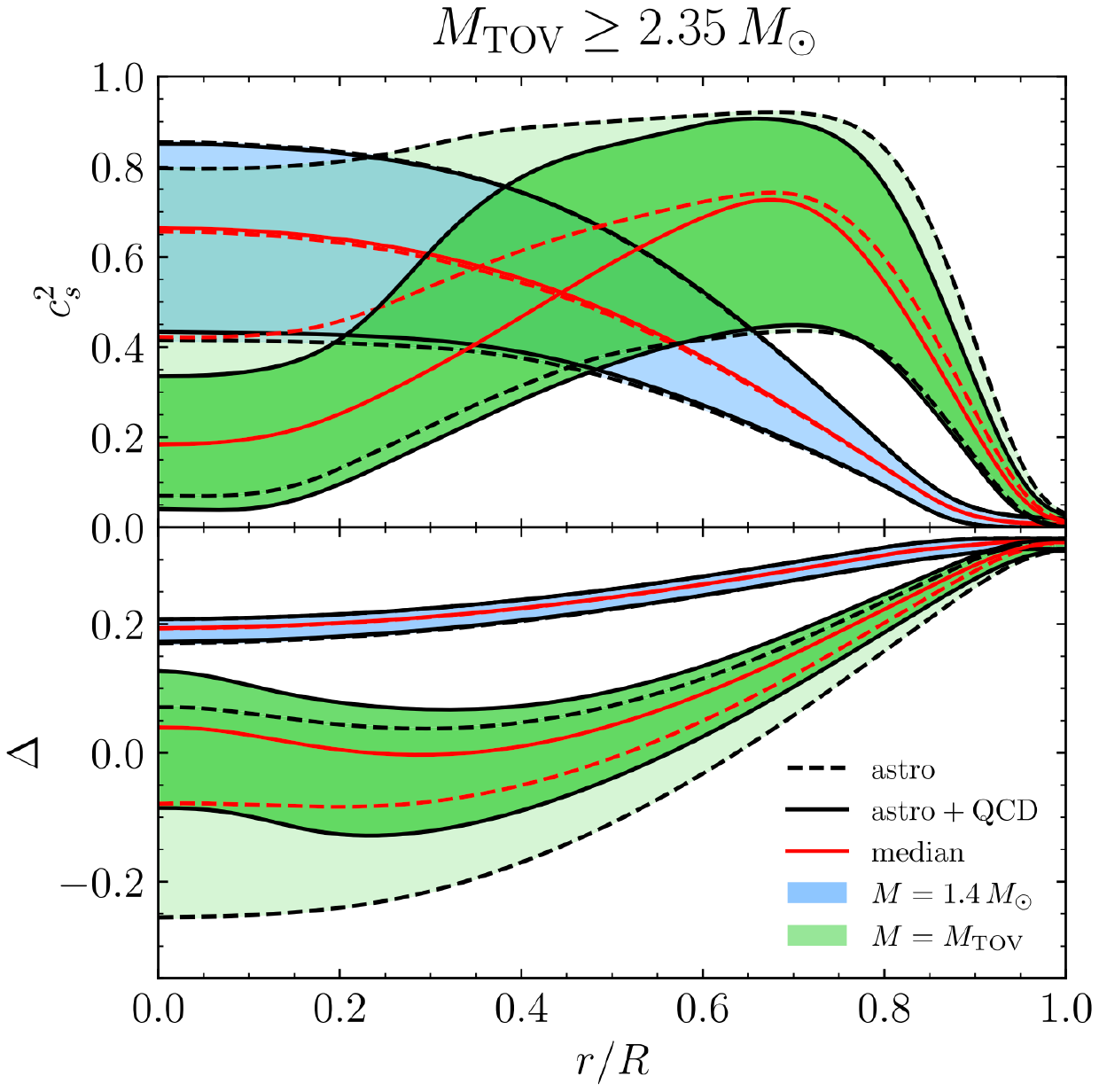}\\[0.25cm]
\includegraphics[height=9.cm]{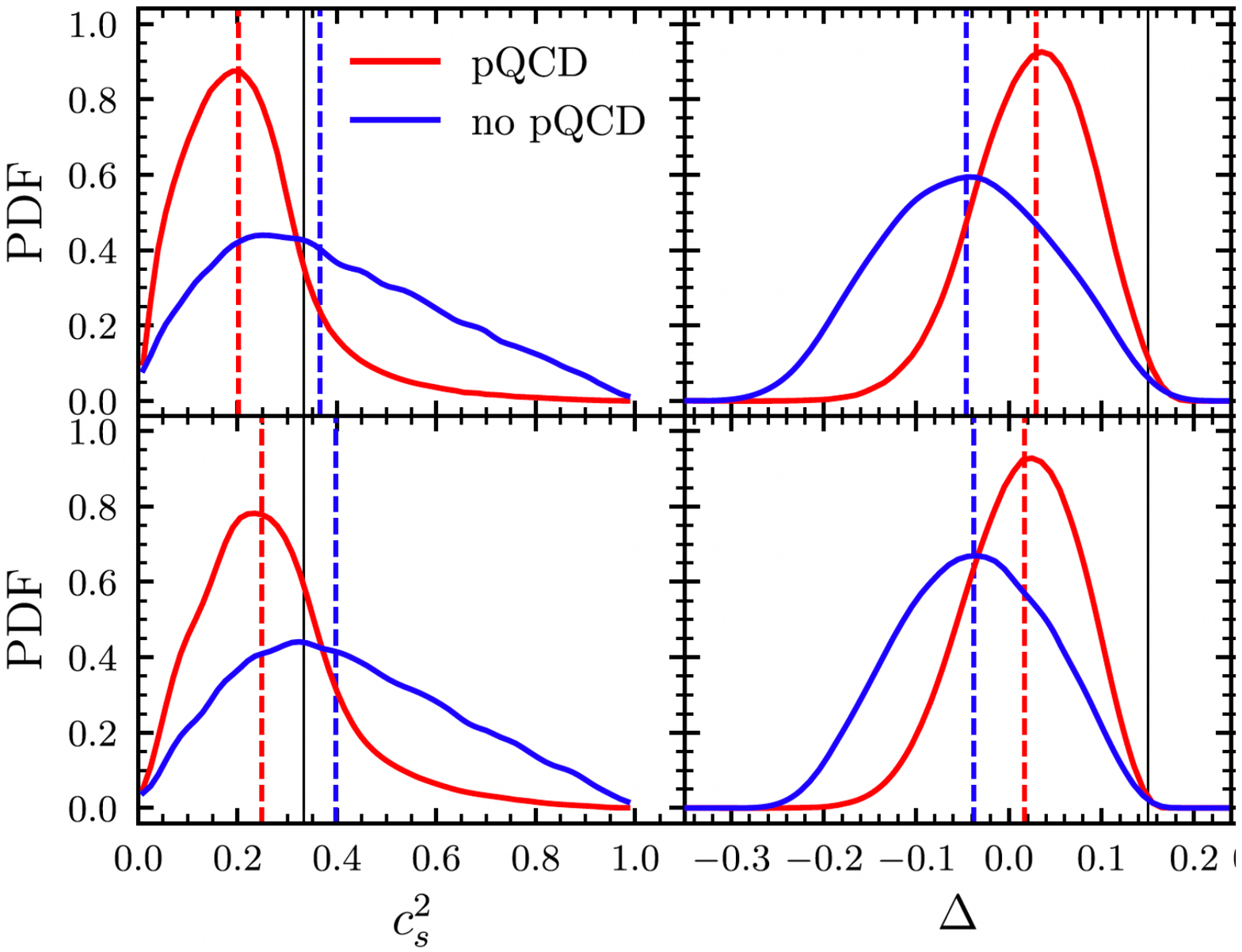}
\caption{\small (Color Online).  Upper panel: radial dependence of $\Delta$ with the constraint $M_{\rm{TOV}}/M_{\odot}\gtrsim2.35$. Figure taken from Ref.\,\cite{Ecker23}. Lower panel: PDF for $\Delta$ with/without considering the pQCD limit at extremely high densities. 
The first (second) line in the lower panel is for non-rotating (Kepler rotating) NSs.
Figure taken from Ref.\,\cite{Mus24}.
}\label{fig_Mus24Delta}
\end{figure}

\begin{figure}[ht!]
\centering
\hspace*{0.3cm}
\includegraphics[height=7.5cm]{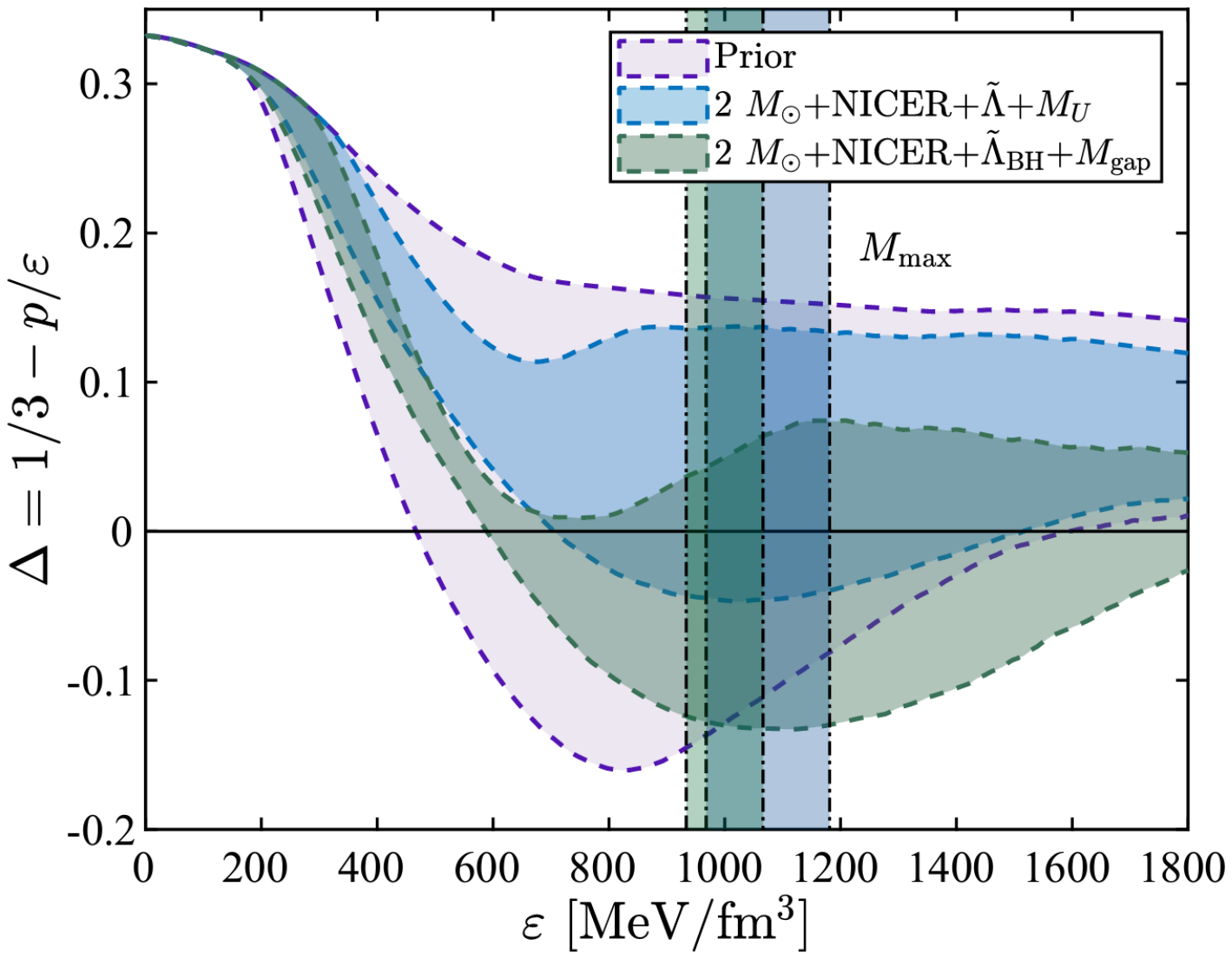}\\[0.5cm]
\includegraphics[height=7.3cm]{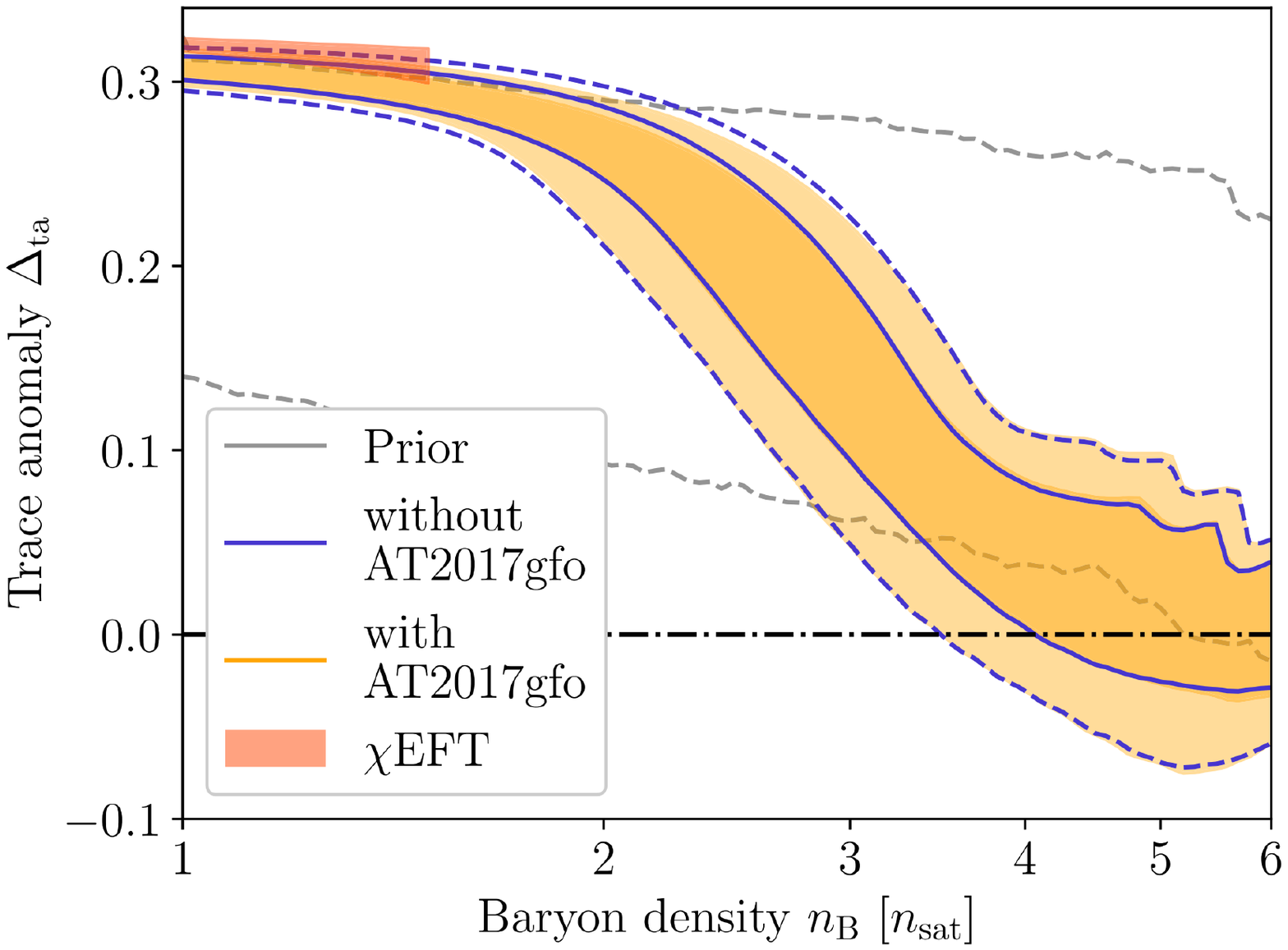}
\caption{\small (Color Online).  Two typical inferences on the energy density (or baryon density) dependence of $\Delta$. Figures taken from Ref.\,\cite{Tak23} (upper panel) and Ref.\,\cite{Pang24} (lower panel).
}\label{fig_Pang24Delta}
\end{figure}

\begin{figure}[ht!]
\centering
\includegraphics[height=8.cm]{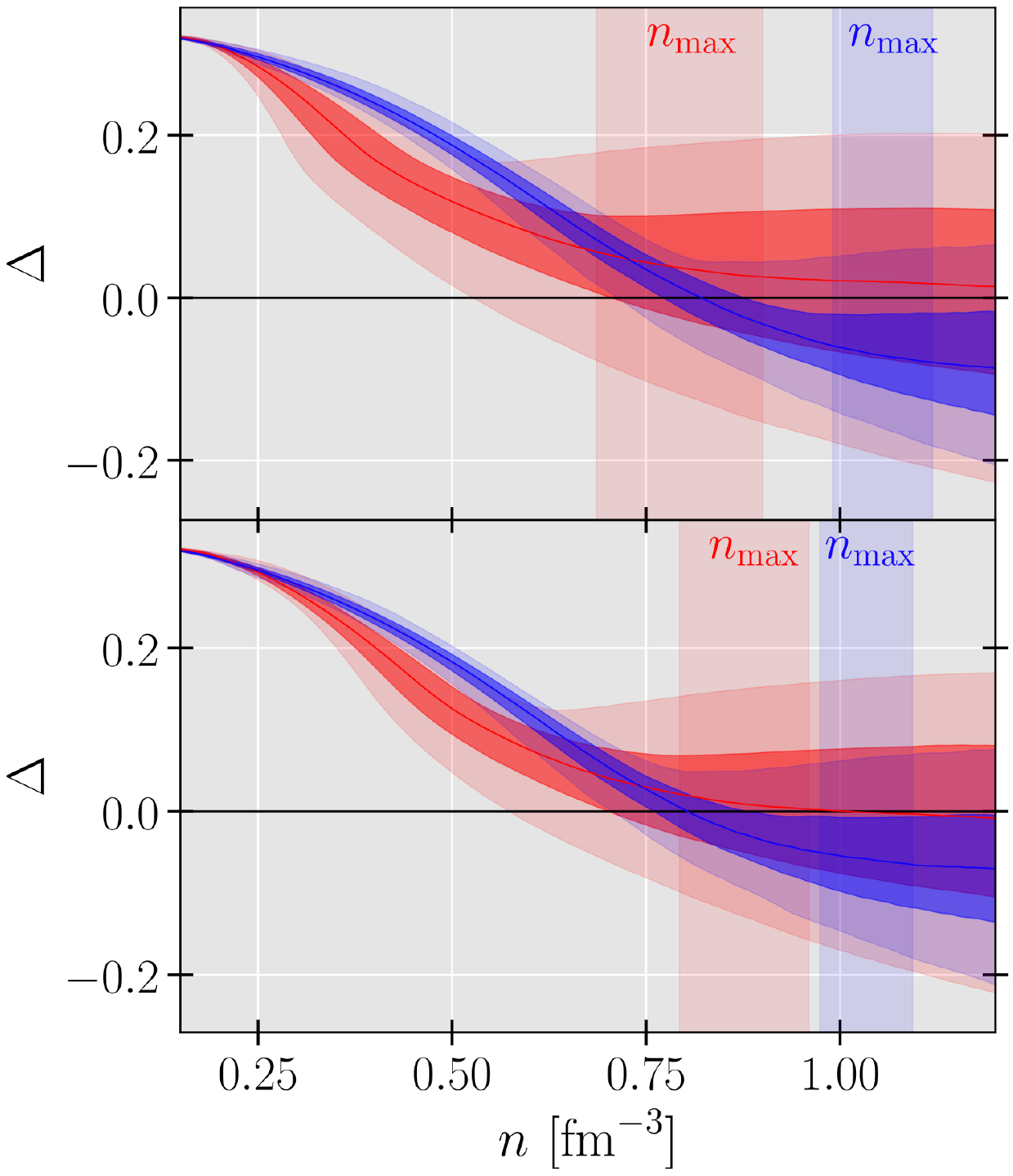}\hspace{1.cm}
\includegraphics[height=8.cm]{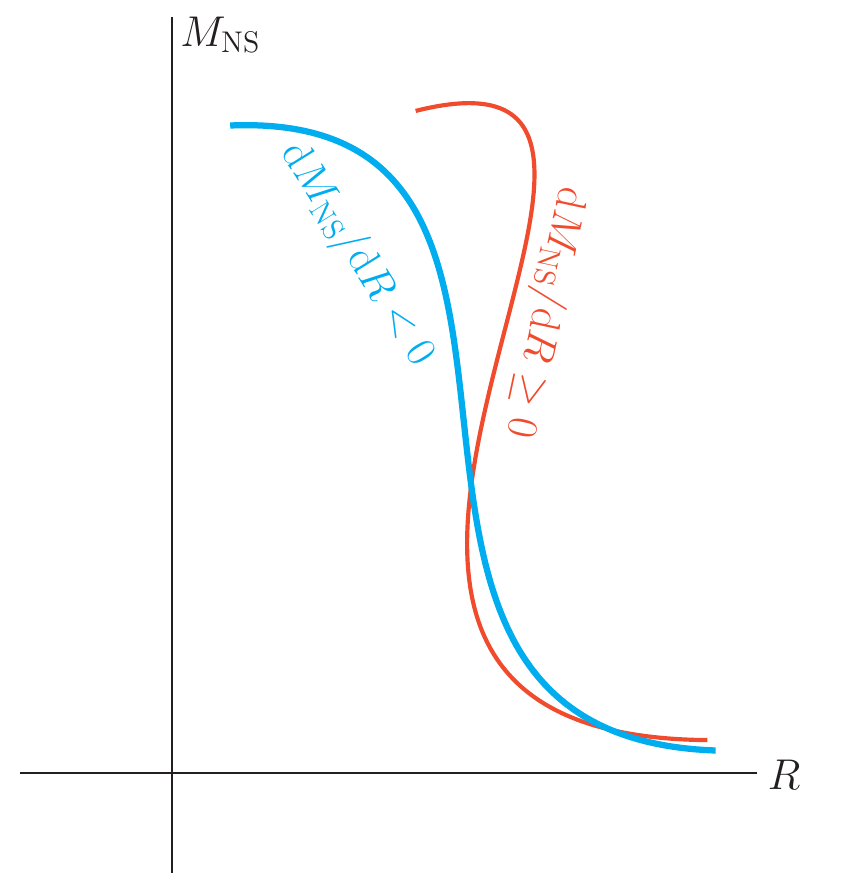}\\[0.5cm]
\hspace{-1.2cm}
\includegraphics[height=10.cm]{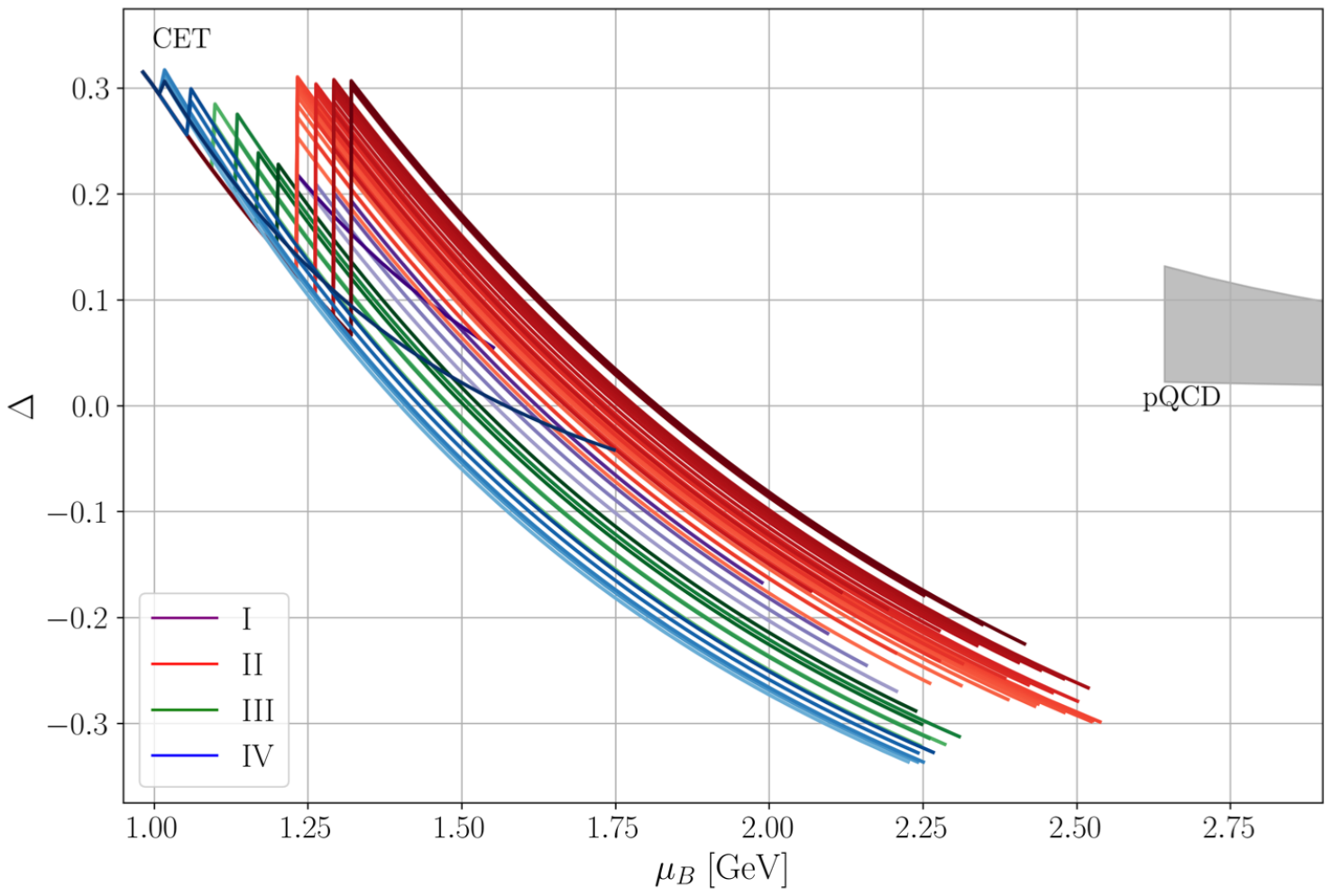}
\caption{\small (Color Online). Upper left panel: density dependence of $\Delta$ inferred under the constraint $\d M_{\rm{NS}}/\d R<0$ for all NS masses (blue) or $\d M_{\rm{NS}}/\d R\geq0$ for a certain mass range (red); inference in the bottom figure with astrophysical constraints. Figure taken from Ref.\,\cite{Ferr24}.
Upper right panel: two types of M-R curves classified by using the derivative $\d M_{\rm{NS}}/\d R$ for NS masses between about $1\,M_{\odot}$ and $M_{\rm{TOV}}$ to help understand the behavior of trace anomaly $\Delta$'s shown in the left panel.
Bottom: The trace anomaly for twin stars satisfying static and dynamic stability conditions. Figure taken from Ref.\,\cite{Jim24}.
}\label{fig_Ferr24Delta}
\end{figure}

The analysis in Ref.\,\cite{Ecker23} using an agnostic EOS showed that $\Delta$ is very close to zero for $M_{\rm{TOV}}\gtrsim2.18\mbox{$\sim$}2.35M_{\odot}$ and may be slightly negative for even more massive NSs (e.g., $\Delta\gtrsim-0.021_{-0.136}^{+0.039}$ for $M_{\rm{TOV}}\gtrsim2.52M_{\odot}$); the radial dependence of $\Delta$ is shown in the upper panel of FIG.\,\ref{fig_Mus24Delta} from which one finds the $\Delta$ for NS at the TOV configuration is much deeper than that in a canonical NS. Moreover, incorporating the pQCD effects ($\Delta_{\rm{pQCD}}\to0$) was found to effectively increase the inference on $\Delta$.
An updated analysis of Ref.\,\cite{Ecker23} was given in Ref.\,\cite{Mus24}, where $\Delta\gtrsim-0.059_{-0.158}^{+0.162}$ or $\Delta\gtrsim 0.019_{-0.129}^{+0.100}$ was obtained under the constraint $M_{\rm{TOV}}\gtrsim2.35M_{\odot}$ without or with considering the pQCD effects, see the lower panel of FIG.\,\ref{fig_Mus24Delta} for the PDFs.
Similarly, if $M_{\rm{TOV}}\gtrsim2.20M_{\odot}$ was required, these two limits become $\Delta\gtrsim-0.046_{-0.166}^{+0.167}$ and $\Delta\gtrsim0.029_{-0.133}^{+0.108}$\,\citep{Mus24}, respectively.
In Ref.\,\cite{Tak23}, the central minimum value of $\Delta$ is found to be about $0.04$ using the NICER data together with the tidal deformability from GW170817,  and a value of $\Delta_{\min}\approx-0.04_{-0.09}^{+0.11}$ was inferred considering additionally the second component of GW190814 as a NS with mass about $2.59M_{\odot}$\,\citep{Abbott2020} using two hadronic EOS models\,\citep{Tak23}, see the upper panel of FIG.\,\ref{fig_Pang24Delta}.
By incorporating the constraints from AT2017gfo\,\citep{Abbott2017-a}, it was found\,\citep{Pang24} that the minimum of $\Delta$ is very close to zero (about $-0.03$ to $0.05$), as shown in the lower panel of FIG.\,\ref{fig_Pang24Delta}.
Using similar low-density nuclear constraints as well as astrophysical data especially including the black widow pulsar PSR J0952-0607\,\citep{Romani22}, Ref.\,\cite{Brandes2023-a} predicted $\Delta\gtrsim-0.086^{+0.07}_{-0.07}$ taken at $\varepsilon\approx1\,\rm{GeV}/\rm{fm}^3$.
Another analysis within the Bayesian framework considering the state-of-the-art theoretical calculations showed that $\Delta\gtrsim-0.01$\,\citep{Ann23} (where $M_{\rm{TOV}}\approx2.27_{-0.11}^{+0.11}M_{\odot}$ is assumed).
Furthermore, by considering the slope and curvature of energy per particle in NSs, Ref.\,\cite{Mar24} showed that $\Delta$ is lower bounded for $M_{\rm{TOV}}$ to be about $-0.02_{-0.03}^{+0.03}$.
In addition, Ref.\,\cite{Cao23} found that the $\Delta$ should be roughly larger than about $-0.04_{-0.09}^{+0.08}$ in self-bound quark stars while that in a normal NS is generally greater than zero.

\begin{figure}[ht!]
\centering
\includegraphics[width=14.cm]{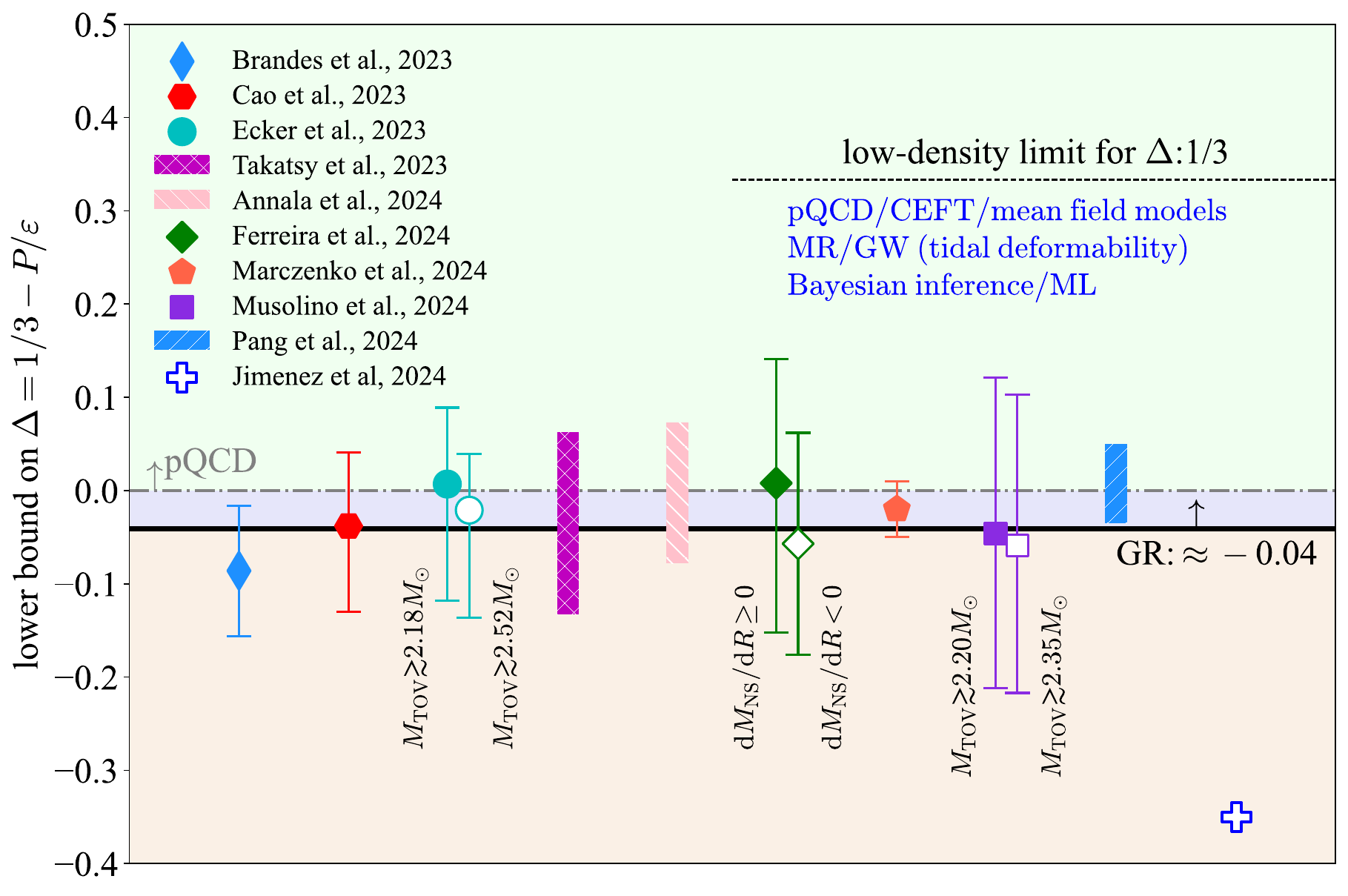}\\[1cm]
\includegraphics[width=12.cm]{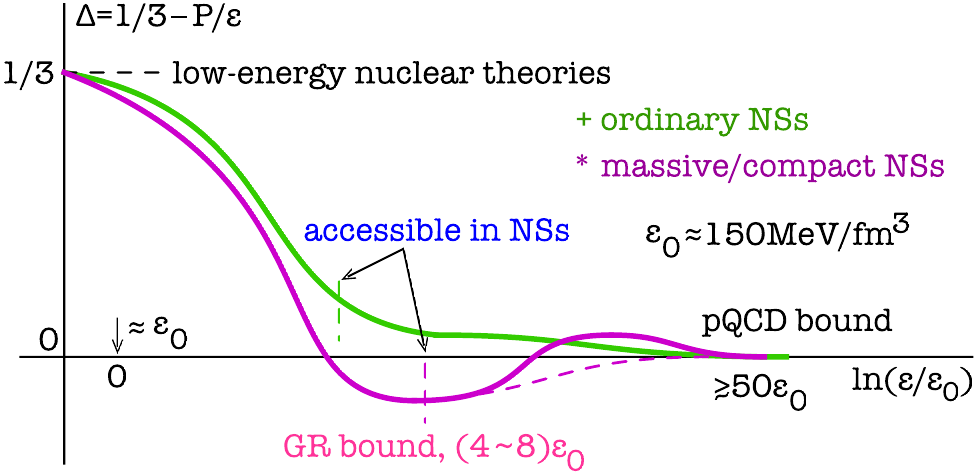}
\caption{\small (Color Online). Upper panel: Summary of current constraints on the lower bound of trace anomaly $\Delta$ in NSs from different analyses with respect to the pQCD (dot-dashed line) and GR (black solid line) predictions. 
See the text for details. Lower panel: Sketch of two imagined patterns for $\Delta=1/3-P/\varepsilon$ in NSs. The $\Delta$ is well constrained around the fiducial density $\varepsilon_0\approx150\,\rm{MeV}/\rm{fm}^3$ by low-energy nuclear theories and is predicted to vanish due to the approximate conformality of the matter at $\varepsilon\gtrsim50\varepsilon_0$ (or equivalently $\rho\gtrsim40\rho_0$) using pQCD theories. The magenta curve is based on the assumption that causality limit is reached in the most massive NS observed where $\varepsilon/\varepsilon_0$ being roughly around 4$\sim$8. Figure taken from Ref.\,\cite{CLZ23-b}.
}\label{fig_TAcomp}
\end{figure}

A very recent study classified the EOSs by using the local and/or global derivative $\d M_{\rm{NS}}/\d R$ of the resulting mass-radius sequences\,\citep{Ferr24}. Limiting the sign of $\d M_{\rm{NS}}/\d R$ to positive on the M-R curve for  NS masses between about $1\,M_{\odot}$ and $M_{\rm{TOV}}$, it was found that $\Delta\gtrsim 0.008_{-0.160}^{+0.133}$\,\citep{Ferr24}. On the other hand, if $\d M_{\rm{NS}}/\d R<0$ is required for all NS masses then $\Delta\gtrsim -0.057_{-0.119}^{+0.119}$ is found, see the upper left  panel of FIG.\,\ref{fig_Ferr24Delta}.
Our understanding on this behavior goes as follows: A negative slope $\d M_{\rm{NS}}/\d R$ along the whole M-R curve with $M_{\rm{NS}}/M_{\odot}\gtrsim1$\,\citep{Ferr24} implies the radius of NS at the TOV configuration is relatively smaller compared with the one with a positive $\d M_{\rm{NS}}/\d R$ on a certain M-R segment, as indicated in the upper right panel of FIG.\,\ref{fig_Ferr24Delta}.
Thus the NS compactness $\xi$ in the former case is relatively larger, which induces a larger $\x$ via Eq.\,(\ref{def-Pic}) and correspondingly a smaller $\Delta$\,\citep{CL24-b}; the smaller radius also implies that the NS is much denser so the maximum baryon density is correspondingly larger\,\citep{Ferr24}.
In another very recent study, the dense matter trace anomaly in twin stars satisfying relevant static and dynamic stability conditions was studied\,\citep{Jim24}. The $\Delta$ was found to be deeply bounded roughly as $\Delta\gtrsim-0.035$\,\citep{Jim24}, as shown in the bottom panel of FIG.\,\ref{fig_Ferr24Delta}.
A deep negative $\Delta$ implies a large $\phi$ or $\x$, so the compactness is correspondingly large according to the relation (\ref{def-Pic}). We notice that the radii obtained in Ref.\,\cite{Jim24} for certain NS masses (e.g., around $2M_{\odot}$) may be small compared with the observational data, e.g., PSR J0740+6620\,\citep{Riley21}.

The above constraints on the lower limit of $\Delta$ (realized in NSs) are summarized in the upper panel of FIG.\,\ref{fig_TAcomp}. Clearly, assuming all results are equally reliable within their individual errors indicated, there is a strong indication that the lower bound of $\Delta$ is negative in NSs. Moreover, except the prediction of Ref.\,\cite{Jim24}, the lower bounds of $\Delta$ from various analyses are very close to the pQCD ($\Delta_{\rm{pQCD}}=0$) or GR limit ($\Delta_{\rm{GR}}\approx-0.04$). It is interesting to note that the $\Delta_{\rm{GR}}$ and $\Delta_{\rm{pQCD}}$ have no inner-relation to our best knowledge currently. However, we speculate that the matter-gravity duality in massive NSs mentioned earlier may be at work here. Certainly, this speculation deserves further studies.

How relevant are the GR or pQCD limit for understanding the trace anomaly $\Delta$ in NSs? 
The $\Delta$ and its energy density dependence are crucial for studying the $s^2$ in NSs\,\citep{Fuji22}. For instance, one can explore whether there would be a peaked structure in the density/radius profile of $s^2$ or not in NSs. 
Sketched in the lower panel of FIG.\,\ref{fig_TAcomp}\,\citep{CLZ23-b} are two imagined $\Delta$ functions versus the reduced energy density $\varepsilon/\varepsilon_0$, here $\varepsilon_0\approx150\,\rm{MeV}/\rm{fm}^3$ around which the low-energy nuclear theories constrain the $\Delta$ quite well. We notice that these two functions are educated guesses certainly with biases. In fact, it has been pointed out that applying a particular EOS in extracting $\Delta$ from observational data may influence the conclusion\,\citep{Mus24}. In the literature, there have been different imaginations/predictions/speculations on how the $\Delta$ at finite energy density may vary and finally reach its pQCD limit of $\Delta=0$ at very large energy densities $\varepsilon\gtrsim50\varepsilon_0\approx7.5\,\rm{GeV}/\rm{fm}^3$\,\citep{Fuji22,Kur10} or equivalently $\rho\gtrsim40\rho_0$. The latter is far larger than the energy density reachable in the most massive NSs reported so far based on our present knowledge. The pQCD limit on $\Delta$ is thus possibly relevant\,\citep{Zhou2024} but not fundamental for explaining the inferred $\phi=P/\varepsilon\gtrsim1/3$ from NS observational data based on various microscopic and/or phenomenological models.
On the other hand, we also have no confirmation in any way that the causality limit is reached in any NS. 
The magenta curve is based on the assumption that the causality limit under GR is reached in the most massive NSs observed so far. Based on most model calculations, in the cores of these NSs the  $\varepsilon/\varepsilon_0$ is roughly around 4$\sim$8. However, if the matter-gravity in massive NSs is indeed at work, we have no reason to expect that the GR limit is reached at an energy density lower than the one where the pQCD is applicable. 

Keeping a positive attitude in our exploration of a completely uncharted area, we make below a few more comments on how the trace anomaly may reach the pQCD limit. As a negative $\Delta$ is unlikely to be observed in ordinary NSs, the evolution of $\Delta$ is probably more like the green curve in the lower panel of Fig.\ \ref{fig_TAcomp}. An (unconventional) exception may come from light but very compact NSs, e.g., a $1.7M_{\odot}$ NS at the TOV configuration with radius about 9.3\,km has its $\Delta_{\rm{c}}\approx-0.02$, since $\varepsilon_{\rm{c}}\approx1.86\,\rm{GeV}/\rm{fm}^3$ together with $P_{\rm{c}}\approx654\,\rm{MeV}/\rm{fm}^3$ should be obtained via the mass and radius scalings of (\ref{Mmax-G}) and (\ref{Rmax-n}) and so $\x=\widehat{P}_{\rm{c}}\approx0.351$.
On the other hand, massive and compact NSs (masses $\gtrsim2M_{\odot}$) are most relevant to observing a negative $\Delta$ (as indicated by the magenta curves) and how it evolves to the pQCD bound, thus revealing more about properties of supradense matter\,\citep{CLZ23-b}.
Interestingly, both the green and magenta curves for the $\Delta$ pattern are closely connected with the density-dependence of the SSS using the trace anomaly decomposition of $s^2$\,\citep{Fuji22} (we do not dive into detailed discussions on these interesting topics in the current review).
Unfortunately, the region with $\varepsilon/\varepsilon_0\gtrsim8$ is largely inaccessible in NSs due to their self-gravitating nature.

\section{Summary and Future Perspectives}\label{SEC_4}
In summary, perturbative analyses of the scaled TOV equations reveal interesting new insights into properties of supradense matter in NS cores without using any input nuclear EOS. In specific, the ratio $\phi=P/\varepsilon$ of pressure $P$ over energy density $\varepsilon$ (the corresponding trace anomaly $\Delta=1/3-\phi$) in NS cores is bounded to be below 0.374 (above $-$0.04) by the causality condition under GR independent of the nuclear EOS.
Moreover, we demonstrate that the NS mass $M_{\rm{NS}}$, radius $R$ and compactness $\xi=M_{\rm{NS}}/R$ strongly correlate with $\Gamma_{\rm{c}}=\varepsilon_{\rm{c}}^{-1/2}\Pi_{\rm{c}}^{3/2}$, $\nu_{\rm{c}}=\varepsilon_{\rm{c}}^{-1/2}\Pi_{\rm{c}}^{1/2}$ and $\Pi_{\rm{c}}=\x/(1+3\x^2+4\x)$ with $\x\equiv\phi_{\rm{c}}=P_{\rm{c}}/\varepsilon_{\rm{c}}$, respectively; therefore observational data on $M_{\rm{NS}}$ and $R$ as well as on $\xi$ via red-shift measurements can directly constrain the central EOS $P_{\rm{c}}=P_{\rm{c}}(\varepsilon_{\rm{c}})$ in a model-independent manner.
Besides the topics we have already investigated\,\citep{CLZ23-a,CLZ23-b,CL24-a,CL24-b}, there are interesting issues to be further explored in this direction. Particularly, we notice:

1.\;The upper limit for $\phi=P/\varepsilon$ near NS cores is obtained by truncating the perturbative expansion of $P$ and $\varepsilon$ to low orders in reduced radius $\hr$. While the results are quite consistent with existing constraints from state-of-the-art simulations/inferences, refinement by including even higher-order $\hr$ terms would be important for studying the radius profile of $\phi$ or $\Delta$ in NSs.
In the Appendix we estimate such an effective correction.

2.\;Ironically, the upper bound $\phi=P/\varepsilon\lesssim0.374$ from GR is very close to that ($P/\varepsilon\lesssim 1/3$) from pQCD at extremely high densities\,\citep{Bjorken83,Kur10,Fuji22}. While we speculated that the well-known matter-gravity duality in massive NSs may be at work, it is currently unclear to us if there is a fundamental connection between them. Efforts in understanding their relations may provide useful hints for developing a unified theory for strong-field gravity and elementary particles in supradense matter. 

\section*{Acknowledgments}
We would like to thank James Lattimer and Zhen Zhang for helpful discussions. This work was supported in part by the U.S. Department of Energy, Office of Science, under Award Number DE-SC0013702, the CUSTIPEN (China-U.S. Theory Institute for Physics with Exotic Nuclei) under the US Department of Energy Grant No. DE-SC0009971.

\appendix
\renewcommand\theequation{A\arabic{equation}}
\setcounter{equation}{0}
\section*{Appendix: Estimate on an Effective Correction to $s_{\rm{c}}^2$}

In this appendix, we estimate an effective correction to $s_{\rm{c}}^2$ given in Eq.\,(\ref{sc2-TOV}) for NSs at the TOV configuration\,\citep{CLZ23-b}.
When writing down $M_{\rm{NS}}$ in Eq.\,(\ref{def-Gc}), we adopt $M_{\rm{NS}}=3^{-1}\widehat{R}^3W$ which only includes the first term in the systematic expansion (\ref{ee-hM});
necessarily we may include higher order terms from (\ref{ee-hM}) in $M_{\rm{NS}}$. As an effective correction we now include $5^{-1}a_2\widehat{R}^5$ from (\ref{ee-hM}) to the NS mass, which modifies Eq.\,(\ref{def-Gc}) as,
\begin{equation}\label{app-1}
    M_{\rm{NS}}\approx\left(\frac{1}{3}\widehat{R}^3+\frac{1}{5}a_2\widehat{R}^5\right)W=\frac{1}{3}\widehat{R}^3W\left(1+\frac{3}{5}a_2\widehat{R}^2\right)
    =\frac{1}{3}\widehat{R}^3W\left(1-\frac{3}{5}\frac{\x}{s_{\rm{c}}^2}\right)
    \sim\Gamma_{\rm{c}}\left(1-\frac{3}{5}\frac{\x}{s_{\rm{c}}^2}\right),
\end{equation}
where $\widehat{R}$ is given by Eq.\,(\ref{def-nc}) through $\x+b_2\widehat{R}^2\approx0$, the coefficient $\Gamma_{\rm{c}}\sim \widehat{R}^3W$ is defined in Eq.\,(\ref{def-Gc}) and the general relation $a_2=b_2/s_{\rm{c}}^2$ is used to write $3a_2\widehat{R}^2/5=-3\x/5s_{\rm{c}}^2$.
The factor ``$1+3a_2\widehat{R}^2/5$'' is actually the averaged reduced energy density $\langle\widehat{\varepsilon}\rangle$ by including the $a_2$-term in $\widehat{\varepsilon}$ of Eq.\,(\ref{ee-heps}), namely $M_{\rm{NS}}/W\approx3^{-1}\widehat{R}^3\langle\widehat{\varepsilon}\rangle$ with
\begin{equation}
\langle\widehat{\varepsilon}\rangle=\left.\int_0^{\widehat{R}}\d\hr\hr^2\widehat{\varepsilon}(\hr)\right/\int_0^{\widehat{R}}\d\hr\hr^2
    =1+\frac{3}{5}a_2\widehat{R}^2,~~
    \widehat{\varepsilon}(\hr)\approx1+a_2\hr^2.
\end{equation}

Moreover, the $s_{\rm{c}}^2$ in Eq.\,(\ref{app-1}) is now not given by Eq.\,(\ref{sc2-TOV}), but should include corrections due to including of the $a_2$-term in $\widehat{\varepsilon}(\hr)$. Generally, we write it as:
\begin{equation}\label{app-2}
    s_{\rm{c}}^2\approx\x\left(1+\frac{1}{3}\frac{1+3\x^2+4\x}{1-3\x^2}\right)\left(1+\kappa_1\x\right)\approx\frac{4}{3}\x+\frac{4}{3}\left(1+\kappa_1\right)\x^2+\mathcal{O}(\x^3),
\end{equation}
where $\kappa_1$ is a coefficient to be determined.
In addition, we have $1-3\x/5s_{\rm{c}}^2\approx(11/20)[1+9(1+\kappa_1)\x/11]$ using the $s_{\rm{c}}^2$ of Eq.\,(\ref{app-2}); taking $\d M_{\rm{NS}}/\d\varepsilon_{\rm{c}}=0$ with $M_{\rm{NS}}$ given by Eq.\,(\ref{app-1}) gives the expression for $s_{\rm{c}}^2$ (which is quite complicated), we then expanding the latter over $\x$ to order $\x^2$ to give
\begin{equation}\label{app-3}
    s_{\rm{c}}^2\approx\frac{4}{3}\x+\frac{1}{11}\left(\frac{38}{3}-2\kappa_1\right)\x^2+\mathcal{O}(\x^3).
\end{equation}
Matching the two expressions (\ref{app-2}) and (\ref{app-3}) at order $\x^2$ gives $\kappa_1=-3/25$.
After that, we determine $\x\lesssim0.381$ via $s_{\rm{c}}^2\leq1$, which is close to and consistent with 0.374 obtained in the main text; and similarly $\Delta\gtrsim-0.048$.
The magnitude of the correction ``$+\kappa_1\x$'' in $s_{\rm{c}}^2$ is smaller than 5\% while the corresponding correction on $\x_+^{\rm{GR}}$ is smaller than 2\%.
In addition, the NS mass now scales as
\begin{equation}
\boxed{
    M_{\rm{NS}}\sim
    \frac{1}{\sqrt{\varepsilon_{\rm{c}}}}\left(\frac{\x}{1+3\x^2+4\x}\right)^{3/2}\cdot\left(1+\frac{18}{25}\x\right).}
\end{equation}

In order to obtain the corrections to $s_{\rm{c}}^2$ more self-consistently and improve the accuracy on $\x_+^{\rm{GR}}$, one may include more terms in the expansion of $\hP$ over $\widehat{R}$ of Eq.\,(\ref{ee-hP}) (i.e., $b_2$-term, $b_4$-term and $b_6$-term, etc.), the expansion of $\hM$ over $\widehat{R}$ of Eq.\,(\ref{ee-hM}) (i.e., $a_2$-term, $a_4$-term, $a_6$-term, etc.) and in the mean while introduce corrections ``$1+\kappa_1\x+\kappa_2\x^2+\kappa_3\x^3+\cdots$'' in $s_{\rm{c}}^2$ as we did in Eq.\,(\ref{app-2}); then determine the coefficients $\kappa_1$, $\kappa_2$ and $\kappa_3$, etc.
The procedure becomes eventually involved as more terms are included.

{\renewcommand*{\bibfont}{\tiny}
\renewcommand{\baselinestretch}{1.1}
\begin{multicols}{2}
\selectfont\fontfamily{lmr}

\end{multicols}
}
\end{document}